\documentclass[conference]{IEEEtran}
\ifCLASSINFOpdf
\else
\fi
\usepackage{amsmath}
\usepackage[utf8]{inputenc}
\usepackage{graphicx}
\usepackage{url}
\usepackage{subfig}
\usepackage{siunitx}
\usepackage{booktabs, tabularx}
\usepackage{cite}

\makeatletter
\renewcommand{\citemid}{\hbox{\@citepunct\hskip 0.13em\relax--\hskip 0.13em\relax}}
\makeatother

\begin{document}
\title{Semantic Communications: the New Paradigm Behind Beyond 5G Technologies}

 \author{Gabriella Fernandes, Hélder Fontes, Rui Campos\\INESC TEC and Faculdade de Engenharia, Universidade do Porto}

\maketitle

\begin{abstract}
Each generation of cellular networks is characterized by its distinct capabilities and innovations, which reflect the significant milestones reached with each new release. The Fifth Generation (5G) of mobile technologies has made substantial progress through the deployment of advanced encoding and modulation techniques, nearly reaching the Shannon physical capacity limit. In light of the requirements of Beyond-5G (B5G) technologies, there is the need of a paradigm shift in the development and conception of communication systems. 

Recent developments in the realm of Artificial Intelligence (AI) have enabled the deployment of tools with high abstraction capabilities, that can be applied to both feature extraction processes and End-to-End (E2E) system optimization tasks. In this context, Semantic Communications has emerged as a novel information transmission system, with AI as one of the core components in its implementation and optimization. This communication paradigm relies on the extraction and transmission of the ``semantic meaning" of the source information using AI techniques, diverging from the conventional systems that primarily focus on ensuring the successful reception of the transmitted bits. By employing a data and semantic-driven approach, Semantic Communications have the potential to yield substantial reductions in communication bandwidth usage, which makes it one of the core principles of the development of B5G networks. 

The purpose of this survey is to provide a comprehensive overview of the fundamental concepts underlying Semantic Communications, including Shannon's Information Theory, classical and modern theories of semantic information, and an examination of the framework and system design of Semantic Communications. Additionally, recent implementations are reviwed, including the analysis of Semantic Communications systems according to the information object transmitted and the objective of the information transmission. Moreover, an in-depth study of prototypes and demonstrations are presented, supporting the viability of the Semantic Communications systems. Finally, some of the most relevant open challenges are detailed, highlighting open research questions to be pursued in Semantic Communications.
\end{abstract}

{\bf Keywords: Semantic Communications, semantic information, B5G, Deep Learning} 

\section{Introduction}
Over the span of approximately three decades, the transition from the First Generation (1G) to the Fifth Generation (5G) of mobile communication has defined profound technological advancements. Targeting the reduction of data reception errors, this evolution traces a shift from the initial analogue signals for voice communication towards the establishment of robust, high-capacity, low-latency communication networks \cite{sc_overview, 6g_beyond_shannon}. With its groundbreaking communication network architecture and key features like Ultra-Reliable Low-Latency Communication (URLLC), massive Machine Type Communications (mMTC), and enhanced Mobile BroadBand (eMBB), 5G is able to integrate sensors, control networks and data management infrastructures \cite{6g_beyond_shannon, ai_enabled, sc_fundamentals, 6g_autonomous_its}. This integration allows 5G to support a wide range of high-demand and stringent requirement services and applications, such as autonomous driving, remote monitoring and controlling, and Internet-of-Things (IoT) systems.

Given the ongoing trends to further integrate data-driven technologies and to increase the levels of automation and intelligence across various domains, it is expected that around 2030 it will be essential to embrace the change towards a Beyond 5G (B5G) network infrastructure \cite{6g_beyond_shannon, ai_enabled}. This transition establishes the foundation for a variety of potential applications of B5G technologies, each requiring increasingly strict specifications. Envisioned applications encompass the development of integrated super-smart cities, the implementation of holographic projections, and intelligent industrial production. Furthermore, B5G is expected to enable the integration of virtual, extended, and augmented reality functionalities, thereby enabling the development of mixed-reality experiences, based on improved user interactions and personalized situations \cite{6g_beyond_shannon, ai_enabled, sc_fundamentals, sc_principles_challenges, metaverse}. Following the tendency of establishing a seamless connection between the digital and physical environments \cite{6g_beyond_shannon, ai_enabled}, the concept of the Metaverse emerges as an advanced stage of the digital transformation \cite{metaverse}. 

B5G applications impose challenges that involve handling the massive volumes of data and a highly interconnected network of nodes. Moreover, these applications demand real-time data exchange to uphold the stringent requirements of Quality of Service (QoS) \cite{sc_fundamentals}. In light of these goals, research efforts have tried to identify some of the requirements and demands to enable these applications \cite{sc_overview, 6g_beyond_shannon, ai_enabled, road_to_6g}, such as:
\begin{enumerate}
    \item \textbf{Traffic capacity:} Starting from a minimum of  \SI{1}{\giga\bit}/s/\SI{}{\square\meter}.
    \item \textbf{Data Rates:} Supporting both downlink and uplink data rates of at least \SI{1}{\tera\bit}/s.
    \item \textbf{Reliability:} A stringent communication reliability of \SI{1e-9}{}.    
    \item \textbf{Low latency:} Achieving ultra-low end-to-end latency, targeting approximately \SI{0.1}{\milli\second}.    
    \item \textbf{Frequency bands:} By making use of broad frequency channels that span up to \SI{3}{\tera\hertz}.    
\end{enumerate}

To address these requirements, it is crucial to explore alternative approaches to the conventional strategy of operating at higher frequencies to allow for broader bandwidths \cite{road_to_6g, ai_native}. Nevertheless, this is limited by the bottleneck imposed by hardware's increasing complexity and energy consumption. Thus, there is a need to implement a communication system that goes beyond Shannon's Information Theory and considers the mathematical limits for correct symbol transmission, regarding the meaning of messages irrelevant to the engineering problem \cite{shannon}. In this context, Semantic Communications emerges as a paradigm based on information extraction to overcome classic communication systems limitations by conveying meaning \cite{sc_fundamentals, road_to_6g}. Hence, its enhanced efficiency, optimized resource allocation and utilization, and improved performance comes as a result from the reduction of transmitted data by focussing on the relevant information for the performed task \cite{6g_beyond_shannon, sc_fundamentals, sc_principles_challenges}.

In \cite{morris}, a three-field division of semiotics, the study of signs, is proposed: \textit{syntactics}, related to the formal connections between symbols; \textit{semantics}, concerning the meaning conveyed by signs; and \textit{pragmatics}, referring to the function of signs and their relationship with their interpreters. Based on the semantic perspective of communication and broadening the analysis presented in Shannon's probabilistic Theory of Communication in \cite{shannon}, Weaver \cite{weaver} introduced the concept of Semantic Communications, highlighting the relevance of the semantic aspect within the field of communication theory. Considering the syntactic, semantic, and pragmatic features of communications \cite{toward_wisdom}, Weaver characterizes the communication problems in three levels:
\begin{enumerate}
    \item \textit{Level A -- Technical problem: } dedicated to ensuring the accurate transmission of symbols, relying on network resources. 
    \item \textit{Level B -- Semantic problem: } pertaining to the accuracy with which the recipient interprets the meaning of symbols transmitted as intended by the sender, meaning the semantic inference of the information.
    \item \textit{Level C -- Effectiveness problem: } referring to the communication efficiency, analysing whether the receiver's interpretation leads to the expected action. In practical terms, this concept manifests in the physical and application realms, where actuation and control take place \cite{6g_beyond_shannon}.
\end{enumerate}

\begin{figure}[htb]
    \centering
    \includegraphics[width=0.5\textwidth]{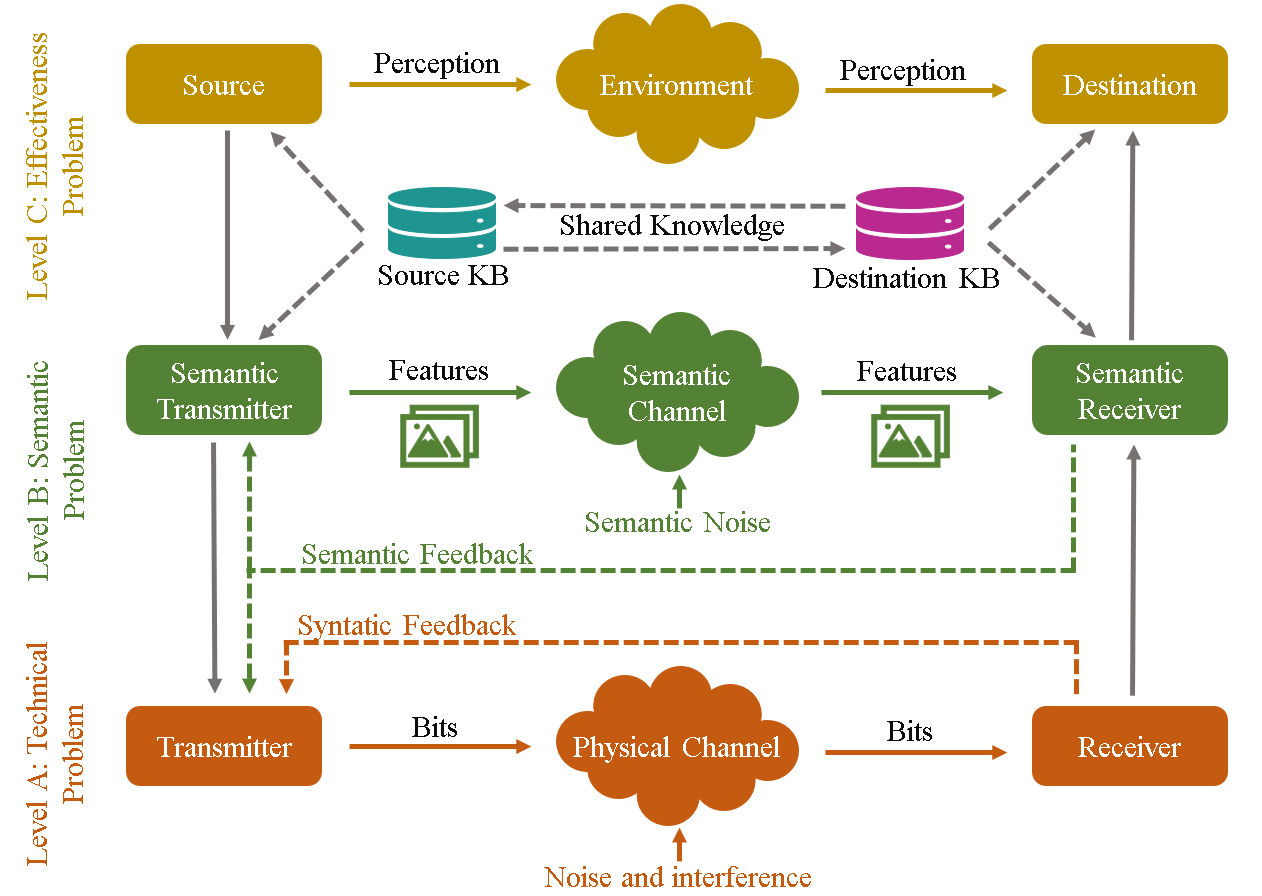}
    \caption{Three level communication system model.}
    \label{fig:3level}
\end{figure}

Figure \ref{fig:3level} elucidates the three-tiered scheme proposed for Semantic Communications, highlighting the interconnections among these levels. Here, \textit{Level C} pertains to source-destination interactions, relying on their Knowledge Bases (KB), \textit{Level B} focuses on the feature extraction and reconstruction, and \textit{Level A} addresses the engineering aspects of communication.

Recent developments of robust Artificial Intelligence (AI) techniques for feature extraction, such as Natural Language Processing and Computer Vision, have led to the creation of novel semantic-based technologies. These advancements have enabled more efficient data transmission by selectively transmitting the task-relevant semantic elements through the physical channel, thereby enhancing transmission efficiency \cite{toward_wisdom}. Moreover, AI's capacity to learn, adapt, and optimize systems is highly regarded in the context of B5G technologies, leading to improvements in network architecture, automation, management, and  performance \cite{ai_native}.

This survey provides a comprehensive overview of the fundamentals of Semantic Communications, introducing the relevant concepts of communications systems, while comparing the traditional and semantic-based communications approaches. Moreover, some of the most relevant recent systems designs are reviewed, highlighting the impact of the information object in the system architecure and performance. Additionally, Semantic Communications systems prototypes are presented, followed by an analysis of the performance of Semantic Communications in real world implementations and extreme scenarios, which provides a unique comprehensive overview in the current real-world deployments. Finally, from the existing works in the field, some of the envisioned challenges and open research questions are explored.

The rest of this paper is organized as follows. Section \ref{sec:1} presents an overview of Classic and Semantic Information Theory. In Section \ref{sec:2}, an overview of the structure of Semantic Communications systems is provided. Moreover, Section \ref{sec:3} presents the impact of Deep Learning in the implementation of Semantic Communications systems. Furthermore, Section \ref{sec:4} presents deployments of Semantic Communications systems. Section \ref{sec:5} reviews some of the open challenges in the context of Semantic Communications, and, finally, in Section \ref{sec:6}, the conclusions drawn from this work are presented.

\section{Paradigm Shift in Communications} \label{sec:1}
This section aims to detail the fundamentals supporting Semantic Communications, as a relevant solution for overcoming the limitations of physical channel capacity. In this context, both an analysis of the evolution from the Classic Information Theory and the various efforts to shape the Semantic Information Theory are presented. 

\subsection{Classic Information Theory}
Shannon's pioneering work in mathematically modelling symbol transmission is agnostic to the semantic content of messages, primarily focusing on the design of systems that enable efficient and accurate message transmission, regardless of their content. In \cite{shannon}, Shannon offers a technical perspective on the communication problem, focusing on the \textit{Level A} of communication. In this context, the fundamental problem of communication is defined as the attempt to reproduce on the receiver side a message that is as similar as possible to the one sent at the source.

\begin{figure}[htb]
    \centering
    \includegraphics[width=0.5\textwidth]{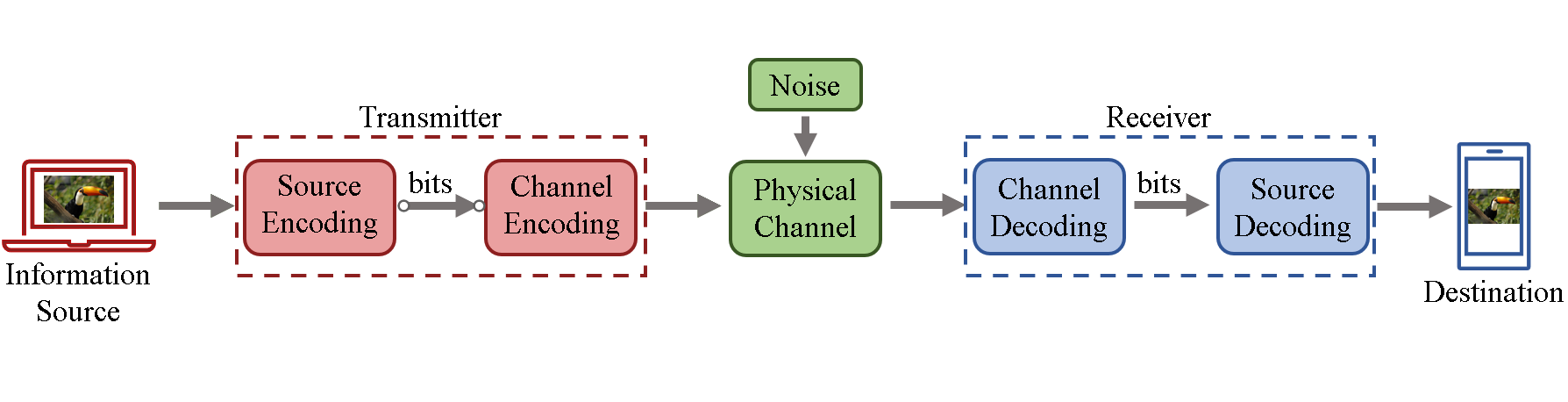}
    \caption{Representation of a traditional communication system.}
    \label{fig:traditional}
\end{figure}

Traditional communication systems are designed according to Shannon's Classic Information Theory and can be represented through block diagrams, with each component operating independently, as represented in Figure \ref{fig:traditional}. Within this context, five key elements characterize a communication system:

\begin{enumerate}
    \item \textit{Information Source:} is responsible for the generating the message to be transmitted;
    \item \textit{Transmitter:} processes the message to be properly transmitted over the channel as a signal. This includes:
    \begin{enumerate}
        \item Source Encoding: removes information redundancy from the raw data, including data compression;
        \item Channel Encoding: adds redundancy and error-correcting codes, to improve reliability.
    \end{enumerate}
    \item \textit{Physical Channel:} refers to the medium through which the signal is transmitted (e.g, optical fibre, coaxial cable). This transmission can be affect by physical channel noise and interference, including Addictive White Gaussian Noise (AWGN) and channel fading;
    \item \textit{Receiver:} reconstructs the message sent from the signal received. This includes:
    \begin{enumerate}
        \item Channel Decoding: removes of redundancy and error-correcting codes;
        \item Source Decoding: reconfigures of data into its original format, possibly executing data decompression.
    \end{enumerate}
    \item \textit{Destination:} refers to the recipient of the message, that is then responsible for the procedures related to \textit{Level C}, including the "meaning" extraction and the actions to be taken accordingly.
\end{enumerate}

One of the core concepts introduced by Shannon was information entropy, which serves as a measure of uncertainty used to quantify the amount of information conveyed in a message. The relevance of this metric extends to its application in forecasting the long-term characteristics of stochastic processes \cite{entropy}. As presented in \cite{shannon}, the entropy for a given source \textit{X} can be inferred from the symbols it generates \(\{x_1, x_2, ..., x_n\}\) with associated probabilities \(\{p(x_i)\}_{i=1}^{n}\), determining the average number of bits per symbol needed for lossless reconstruction \cite{sc_principles_challenges}, as given in Equation \ref{eq:eq1}.

\begin{equation}\label{eq:eq1}
    H(X) = -\sum_{i=1}^{n}p_i \log_{2}(p_i)
\end{equation}

The concept of channel capacity defines an upper bound on the data rate for reliable communication over a noisy channel. In the context of information theory, the uncertainty associated with the received signal, quantified as the conditional entropy of the received signal given the transmitted signal \(H_y(x)\), is a direct reflection of the information lost due to interference and noise. Thus, the transmission rate \textit{R} of the source is defined as presented in Equation \ref{eq:eq2}, where \(H(x)\) denotes the amount of information sent.

\begin{equation}\label{eq:eq2}
    R = H(x) - H_y(x)
\end{equation}

Furthermore, it is possible to infer that the channel capacity \textit{C} for a noisy channel corresponds to the maximum possible transmission rate, as given in Equation \ref{eq:eq3}. Given that the mutual information measures the shared information between \textit{X} and \textit{Y}, the maximum mutual information achievable over all possible input probability distributions \(p(x)\) can be seen as a fundamental measure of how much information can be reliably transmitted over a noisy channel, as represented in \ref{eq:eq4}.

\begin{equation}\label{eq:eq3}
    C = max(H(x) - H_y(x)) 
\end{equation}

\begin{equation} \label{eq:eq4}
    C = \underset{p(x)}{max} I(X;Y)
\end{equation}

Moreover, some of the metrics employed to assess the performance of traditional communication systems include the Signal-to-Noise Ratio (SNR), the Bit Error Rate (BER), and the Symbol Error Rate (SER). The SNR refers to the ratio between the signal strength and the power of the noise affecting the transmission, with higher values indicating higher transmission quality. In turn, the BER and SER measure, respectively, the numbers of incorrect bits and symbols received at the destination, with lower values indicating reduced errors within the channel.

These early developments in the Information Theory significantly contributed to the evolution of conventional communication systems. The Classical Information Theory has facilitated a wide spectrum of applications and has served as a foundational element in the implementation of state-of-the-art technologies, including those employed in the development of 5G mobile communication systems. 

\subsection{Classic Semantic Information Theory}
In an effort to build upon Shannon's study by exploring the semantic and effectiveness levels of communication, Weaver \cite{weaver} provided a pioneer analysis of the role of semantics in the communication problem. Weaver argued that while Shannon's theory was originally developed to address technical communication problems, it exhibited sufficient generality to contribute to the modelling of \textit{Levels B} and \textit{C}. 

In \cite{carnap_bar}, Bar-Hillel and Carnap introduce the Theory of Semantic Information, considering the semantic information as the meaning of the messages. Based on propositional logic, one of the main contributions of this work is the estimation of the amount of information, as presented in Equation \ref{eq:eq5}. Here, the amount of information \emph{inf} conveyed by a proposition \emph{i} is expressed through the degree of confirmation \emph{cont} of the hypothesis \emph{i} by the evidence \emph{e}.

\begin{equation} \label{eq:eq5}
    inf(i) = -\log_{2}(1 - cont(i))
\end{equation}

Nevertheless, the Theory of Semantic Information introduces a paradox by implying that contradictions contain the highest amount of semantic information. Motivated by this problem, Floridi \cite{floridi} developed the Theory of Strongly Semantic Information, relying on truth-values and situational logic instead of the probability distributions used in \cite{carnap_bar}. Moreover, D'Alfonso \cite{dalfonso} proposes the quantification of the amount of semantic information to rely on the definition of truthlikeness, supporting a broader range of possible statements \cite{toward_wisdom, towards_sc}.

\subsection{Modern Information Theory}
A more recent assessment of the research conducted by Floridi and D'Alfonso is presented in \cite{towards_sc}. This analysis points out a limitation in their information quantification methods, as they rely on relative measures, demanding a reference statement for context. Thus, this work studies the semantic parallels of Shannon's Information Theory, providing definitions of semantic capacity, semantic noise, and a semantic channel capacity. With a similar approach to \cite{carnap_bar}, focussing also in propositional logic, it was proposed that the probability of a message \emph{x}, for which \(W_x = \{w \in W | w \models x\}\) is a set of its models, is given by Equation \ref{eq:eq6}. Here, it is considered that \(\mu\) refers to a probability measure in the absence of any background knowledge constraints, meaning that \(\sum_{w\in W}\mu(w) = 1\).

\begin{equation} \label{eq:eq6}
    m(x) = \frac{\mu(W_x)}{W} = \frac{\sum_{w\in W, w \models x}\mu(w)}{\sum_{w\in W}\mu(w)}
\end{equation}

The semantic entropy of a message \emph{x} is then described by Equation \ref{eq:eq7}.
\begin{equation} \label{eq:eq7}
    H(x) = - log_{2}(m(x))
\end{equation}

Additionally, Bao et. al also models the semantic entropy for the presence of background knowledge between the source and receiver, indicating that previously shared information allows source compression without semantic information loss. Moreover, they consider that semantic noise comes from mismatched interpretations regarding true and false information between the source and receiver. Additionally, a theorem for the semantic channel capacity \(C_s\) for discrete memoryless channels was described as in Equation~\ref{eq:eq8}. 

\begin{equation} \label{eq:eq8}
    C_s = \underset{p(Z|X)}{sup}\{I(X;V) - H(Z|X) + \overline{H_s(V)}\}
\end{equation}

As communication paradigms shift towards having semantic information as the core of communication systems, the study introduced in \cite{kountouris_pappas} introduces a purpose-driven integration of data generation, information transmission, and utilization. This proposed system strives to facilitate real-time data transmission, ensuring accurate reception and action by the recipient. In this context, the semantics of information serves as a metric for its usefulness in achieving task goals. Through the definition of semantic information at various levels of granularity, it increases cohesion within the end-to-end communication system, allowing its optimization for the given task.  

In an attempt to improve the representation of semantic symbols to correct interpretation of the transmitted message, the work developed in \cite{learning_semantics}. The proposed architecture relies on the theoretical characterization of different aspects of the system model for Semantic Communications, including semantic source and channel coding, semantic decoder, and semantic channel and noise. For the semantic source and channel coding, taking into account the effects of semantic redundancy and ambiguity, the complexity of the encoder for the semantic compression \emph{I(X;M)} should be minimized. Considering the semantic decoder, one of the system's objectives is to minimize semantic distortion, measured as the expected Kullback Leiber (KL), represented as \emph{EKL}, divergence between intended and decoded messages. Regarding the semantic channel and possible interferences affecting the message, the solution encompasses maximizing the mutual information \(I(X;Y)\) between the two sides of the channel. Putting together all the aforementioned requirements and goals, the proposed solution includes an optimization process guided by the minimization of the objective function defined in Equation \ref{eq:eq11}, where parameters $\alpha$ and $\beta$ modulate trade-offs.

\begin{equation} \label{eq:eq11}
    L = I(X;M) - (1 + \alpha)I(X;Y) + \beta EKL(X;Y)
\end{equation}

Given that previous approaches to model Semantic Communications are strictly task-specific, the work developed in \cite{theory_sc} presents a universal approach to measure semantic entropy and semantic channel coding, being the first work to actually prove that Semantic Communications is able to exceed Shannon's physical channel limit. The proposed semantic entropy, \(H_s(x)\), is based on the uncertainty associated with the interpretation of the transmitted symbols given the knowledge base, as described in Equation \ref{eq:eq9}, with \(p(s_x)\) corresponding to the probability distribution of the semantic interpretation of the value \emph{X}, \(S_x\). In turn, the channel capacity for Semantic Communications systems \(C_s\) is described by Equation \ref{eq:eq10}, considering the discrete memoryless channel \(p(y|x)\) and the mapping \(f_s : \left[ 1:2^{nR} \right]\longrightarrow \left[ 1:2^{\alpha nR} \right]\) associating the source message \emph{W} to \emph{S}, the inherent information of \emph{W}.

\begin{equation}\label{eq:eq9}
    H_s (X) = \sum_{s_x}p(s_x)\log_{2}p(s_x)
\end{equation}

\begin{equation}\label{eq:eq10}
    C_s = \underset{p(x)}{max}{\frac{I(x;Y)}{\alpha}}
\end{equation}

In light of its early development stage, it is important to emphasize that there is no universally accepted approach for quantifying semantic information or to characterize semantic channel capacity. However, the review provided here offers a comprehensive overview of the potential methods for modelling Semantic Communications systems.

\section{Semantic Communications Systems}\label{sec:2}
Traditional communication systems aim to ensure the accurate decoding of transmitted symbols and strive to achieve the highest possible transmission rates. This process requires no intelligent elements, and the semantic information behind the transmitted message is irrelevant \cite{sc_overview}. In contrast, Semantic Communications systems focus on the correct interpretation of the message at the receiver, where the precise transmission of bits is less critical as long as the semantic information is conveyed \cite{6g_beyond_shannon}. Thus, by selectively transmitting only contextually relevant data to the receiver \cite{sc_principles_challenges}, the system achieves substantial resource efficiency gains, leading to bandwidth and power consumption reduction.

In Semantic Communications systems, the semantic information refers to the task-relevant data to be transmitted. Hence, these systems include the existence of highly intelligent agents in the source and destination, whose role is to perform tasks of semantic encoding and decoding. After the message's semantic information interpretation, the effectiveness of the Semantic Communications system is related to the posterior execution of the expected actions by the receiver \cite{kountouris_pappas}.

\begin{figure}[h]
    \centering
    \includegraphics[width=0.5\textwidth]{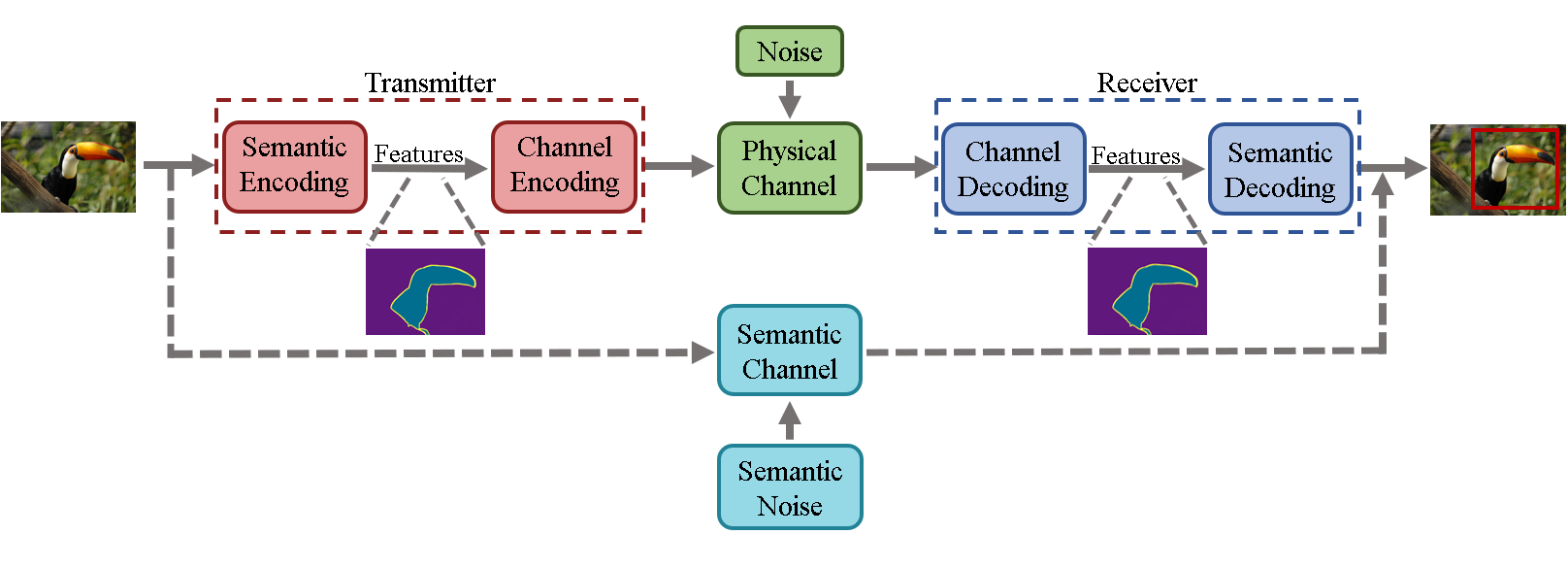}
    \caption{Representation of a Semantic Communications system.}
    \label{fig:sc}
\end{figure}

As already envisioned by Weaver in \cite{weaver}, given the generality of the block diagram presented in Shannon's Theory of Communication, depicted in Figure \ref{fig:traditional},  the representation of Semantic Communications systems would simply encompass incorporating the semantic aspects of the system. As exemplified by the task of image recognition in Figure \ref{fig:sc}, a Semantic Communications system further includes a semantic channel, semantic noise, a semantic encoder and a semantic decoder. Relying on the KB, the semantic encoder is responsible not only for the data compression, but also for the feature extraction of the message to improve the precision of the information representation \cite{sc_overview, survey_on_sc}. The extracted semantic information then undergoes traditional channel coding and decoding processes to be transmitted through the physical channel \cite{survey_on_sc}. In addition, at the receiver, the semantic decoder aims to process the received semantic information to reconstruct the original message. The semantic channel represents an artificial channel connecting the source and destination \cite{sc_overview}, through which the semantic information is transmitted, under the interference of semantic noise.

Additional elements for the Semantic Communications system are referred in \cite{kountouris_pappas} and \cite{signal_processing}. In \cite{signal_processing}, a goal-oriented universal signal processing approach is presented. With a graph-based framework, the semantic signal processing can be tailored for a specific application. In turn, in \cite{kountouris_pappas}, the steps to deploy a semantic-aware network to meet reliability and time constraints of real-time applications are proposed, including semantic filtering through sampling, semantic preprocessing, semantic reception and semantic control. 

Nevertheless, it is relevant to refer that there is no universal model representing Semantic Communications systems as of now. In \cite{popovski}, the integration of semantic components into the conceptual Open Systems Interconnection (OSI) model was proposed to prevent increasing protocol stack complexity. Their recommendation is to incorporate a semantic sub-layer within the application layer. On the other hand, in \cite{toward_wisdom}, a comprehensive Semantic Communications network architecture is proposed, interconnecting the different protocol layers through a semantic intelligence plane. 

\subsection{Knowledge Base}
As a knowledge-based system \cite{6g_beyond_shannon}, Semantic Communications frameworks rely on intelligent agents to interpret information based on their KB. The KB, in this context, refers to the background knowledge to which the agent has access, gathered from the information to which it has been exposed from knowledge sharing and training \cite{6g_beyond_shannon}. Such a framework resembles human interactions, where the way one interprets information is directly related to their previous experiences and the knowledge acquired. Moreover, similarly to the human behaviour, intelligent agents can have different understandings of the same information, which may cause semantic mismatches in the message transmission. 

Nevertheless, although dependent on the background knowledge, one of the main advantages of Semantic Communications systems lies in the ability to transmit information effectively, even when the exact message is not perfectly reconstructed. For example, when the source conveys the message ``The house is there," and the destination receives ``The building is there," no semantic information is lost. This resilience is ensured by the system's capacity to represent information in multiple forms.

\subsection{Semantic Noise}
In traditional communications, the quality of the received signal is affected by physical channel noise and interference, as depicted in Figure \ref{fig:traditional}. In Semantic Communications, in addition to the physical noise, the transmitted information is also under the influence of semantic noise. The occurrence of semantic errors can be attributed to the disruptions introduced by semantic noise during message exchange, which may impact the receiver's interpretation of the messages \cite{sc_fundamentals, deepsc, sc_overview}. These misinterpretations can, in turn, affect the overall effectiveness of the communication system and influence decision-making processes \cite{sc_principles_challenges}.

Disturbances in the interpretation process mostly come as a result of possible semantic ambiguities in the information sent. This arises from possible discrepancies between the KBs of the sender and receiver, which are essential in the process of feature extraction and message reconstruction. This challenge can be illustrated by polysemic expressions, which can lead to misinterpretations even when no syntactic errors occur during physical transmission. For instance, the word ``Apple" could refer to both the fruit and the technology brand. To avoid semantic ambiguities, it is possible to implement real-time sharing of the local knowledge or to resort to the same KB for both the source and destination \cite{sc_fundamentals, sc_overview}, so that the semantic noise does not impair the message interpretation.

\subsection{Performance Metrics}
In the context of prioritizing the transmission of semantic information over the precision of bit-level accuracy, conventional performance metrics like SER and BER rate are no longer relevant or suitable. Thus, the performance of Semantic Communications systems are usually evaluated through the semantic similarity between the information extracted at the transmitter and the decoded information at the receiver \cite{survey_on_sc}. However, due to the significant variations among AI tasks, a universally applicable metric that guarantees data agnosticism and can be utilized for diverse semantic information transmission tasks has not yet been established. 

In the context of text transmission, it is possible to rely on the Bilingual Evaluation Understands (BLUE) score \cite{bleu} and the BERT SCORE \cite{bertscore}. Initially designed as a metric for the assessment of machine translation tasks, the BLEU score can be adapted to other natural language general tasks, making it a suitable metric for text generation in Semantic Communications systems. The BLEU score is then able to measure the similarity between the transmitted and received sentences by comparing the n-grams, which refer to a series of \textit{n} words, in the two sentences. This process yields a value between 0 and 1, with values closer to 1 referring to higher similarity between sentences.

Nevertheless, an accurate evaluation of the performance of a Semantic Communications system for text transmission needs to go beyond a word matching criteria. It should include an in-depth analysis of the semantic information embedded within the transmitted and received sentences, being sensitive to linguistic complexities such as synonyms and polysemy \cite{sc_overview, robust}. In this context, BERT SCORE emerges as a metric capable of assessing semantic similarity by comparing the contextual sentence embeddings, thus capturing the underlying semantic information of the sentences.


Regarding audio data, the performance is measured according to the intelligibility of the decoded speech \cite{sc_fundamentals}. For this task, the Mel-Cepstral Distance (MCD) \cite{mcd}, the Signal-to-Distortion Ratio (SDR) \cite{sdr}, the Perceptual Evaluation of Speech Distortion (PESQ) \cite{deepsc_s}, and the Perceptual Objective Listening Quality Assessment (POLQA) \cite{polqa} can be used to assess the quality of the reconstructed audio signal.

MCD, a perception-based metric for evaluating speech reconstruction, resorts to Cepstral Distance (CD) on the Mel frequency scale to model the spectral characteristics of two audio signals, quantifying their similarity. It serves as a measure of speech reconstruction quality, with a smaller MCD indicating a closer match to the original spectrum \cite{sc_speech}. Moreover, the SDR measures audio quality by implementing a raw speech vector matching analysis between the transmitted and receive speech signals \cite{sc_overview}, with higher SDR values representing higher similarity. In turn, PESQ is an accurate measure for speech quality standardized by the International Telecommunication Union (ITU-T) that considers the effect of a wide range of conditions, including background noise and analogue filtering \cite{pesq}. POLQA, an evolution of the PESQ, aims to quantify the perceived quality of speech signals, broadening its scope to encompass various distortions, including reverberations, time stretching, compression, and playback volume. This next-generation perceptual objective speech quality measurement algorithm, also standardized by ITU-T \cite{polqa}, has demonstrated substantial improvements over its predecessor. While these metrics are highly accurate for assessing audio signal reconstruction quality, it is important to highlight that they rely exclusively on acoustic and spectral signal characteristics and do not take into account the semantic content of the speech.

When it comes to the transmission of images, some of the most relevant traditional methods to quantify the similarity between the transmitted and received images are the Peak Signal-to-Noise Ratio (PSNR) and the Structural Similarity Index (SSIM). PSNR quantifies image fidelity by comparing the maximum SNR between the two images, with higher PSNR values indicating greater similarity. Thus, PSNR is essential in indicating the fidelity of the reconstructed image, recognizing distortions \cite{nextg}. In contrast, SSIM goes beyond pixel values, considering structural elements, luminance, and contrast changes in the image. 

Additionally, as an extension of SSIM, the Multi-Scale Structural Similarity (MS-SSIM) \cite{mssim} incorporates multiple scales, enabling abstracting texture features, preserved across different scales \cite{nextg}. Moreover, the Visual Information Fidelity (VIF) \cite{vif} comes as a visual fidelity measure based on natural scene statistics, comparing images through their statistical correlation, which allows determining whether the origanl image's details are present in the reconstructed image \cite{nextg}.

Nevertheless, it is relevant to note that both PSNR and SSIM are shallow functions that, meaning that they have a limited representation power and are not able to extract semantic information from the images \cite{sc_fundamentals, sc_principles_challenges}. An effective strategy to address this limitation is to resort to DL to extract semantic information. This process is enabled by the use of the weights of the hidden layers of convolutional networks trained for a specific task \cite{sc_fundamentals}. Furthermore, the work carried out in \cite{deep_features} supports that the perceptual similarity is a characteristic of deep visual representations, meaning that the semantic information of images can be extracted from different DL architectures and varying levels of supervision.

This is exemplified by the vast use of Visual Geometry Group (VGG) \cite{vgg} DL architecture trained on ImageNet for the quantification of the similarity of two images and employed for various image comparison analysis, including for the calculation of loss functions for adversarial networks. Moreover, the Learned Perceptual Image Patch Similarity (LPIPS) \cite{lpips} uses the features from pre-trained NNs as input to a smaller network and measures their similarity according to the L2 distance between the deep features \cite{severo2023unreasonable}.

\section{Deep Learning in Semantic Communications Systems} \label{sec:3}
Machine Learning (ML) is an approach for building intelligent systems that can operate in complex real-world contexts and improve with experience and data  \cite{deep_learning, dl_wireless}. Moreover, these ML methods are categorized based on the nature of the learning process: supervised learning involves establishing associations within datasets by mapping features to corresponding labels, unsupervised learning enables algorithms to infer dataset properties from features, and reinforcement learning is based on algorithms learning through interactions with the environment \cite{dl_phy_comm}. Furthermore, one of the most important aspects of ML is its capacity to execute optimization tasks. With the ability to learn patterns and relationships from data, ML aims to minimize a specified loss function tailored to achieve a certain goal.

In this context, DL  arises as a subcategory of ML characterized by a greater degree of abstraction in comparison to conventional ML methodologies. DL is able to provide a heightened level of abstraction to encapsulate complex concepts through simpler representations \cite{deep_learning}. Thus, one of the most relevant characteristics of DL systems is to generate approximations to virtually any function, establishing themselves as universal function approximators \cite{dl_wireless}. Additionally, DL demonstrates improved versatility and effectiveness given its ability to efficiently and parallelly handle large datasets \cite{dl_wireless}.

\begin{figure}[ht]
    \centering
    \includegraphics[width=0.5\textwidth]{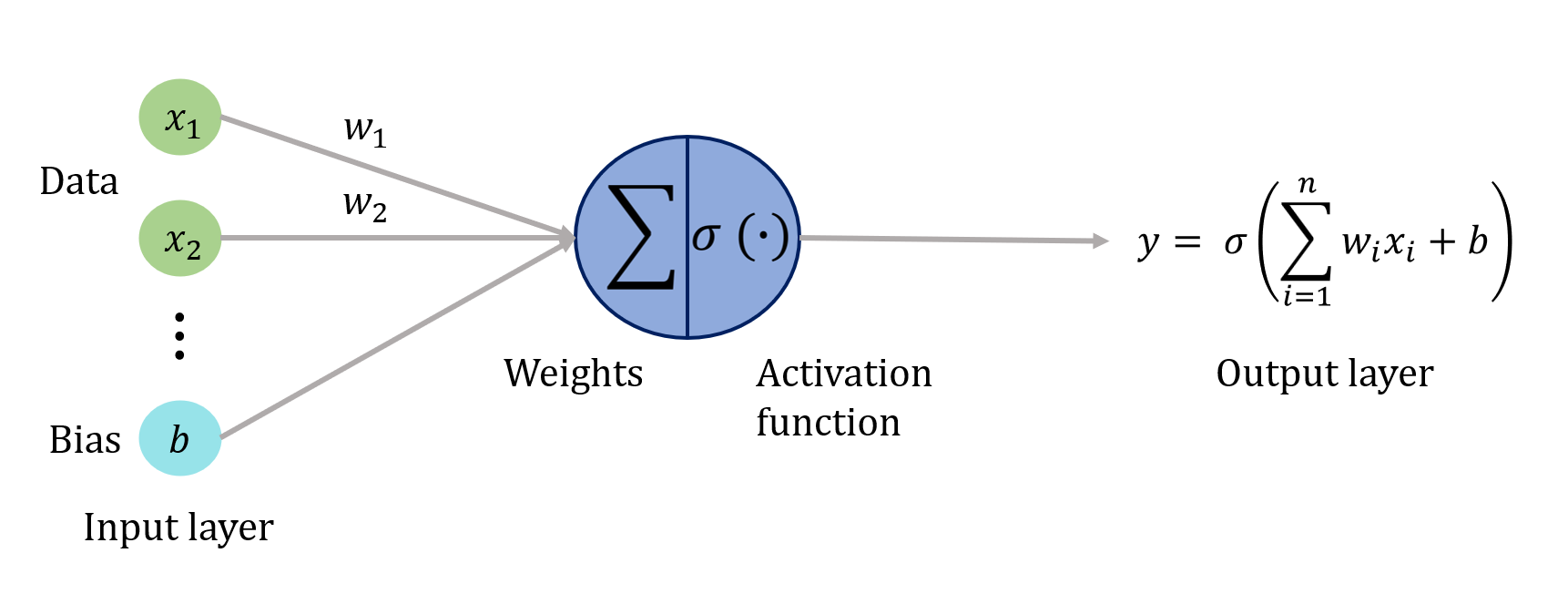}
    \caption{Single neuron.}
    \label{fig:perceptron}
\end{figure}

In an attempt to mimic the human brain \cite{dl_wireless}, DL models have the neuron as its fundamental computational units. As illustrated in Figure \ref{fig:perceptron}, these neurons can be conceptualized as activation functions \(\sigma\left( \bullet  \right)\) that take in the weighted sum of raw input data along with the bias, producing an output \emph{y}. Thus, a Neural Network (NN) is obtained from interconnection of diverse neurons in a layered architecture. 

\begin{figure}[ht]
    \centering
    \includegraphics[width=0.5\textwidth]{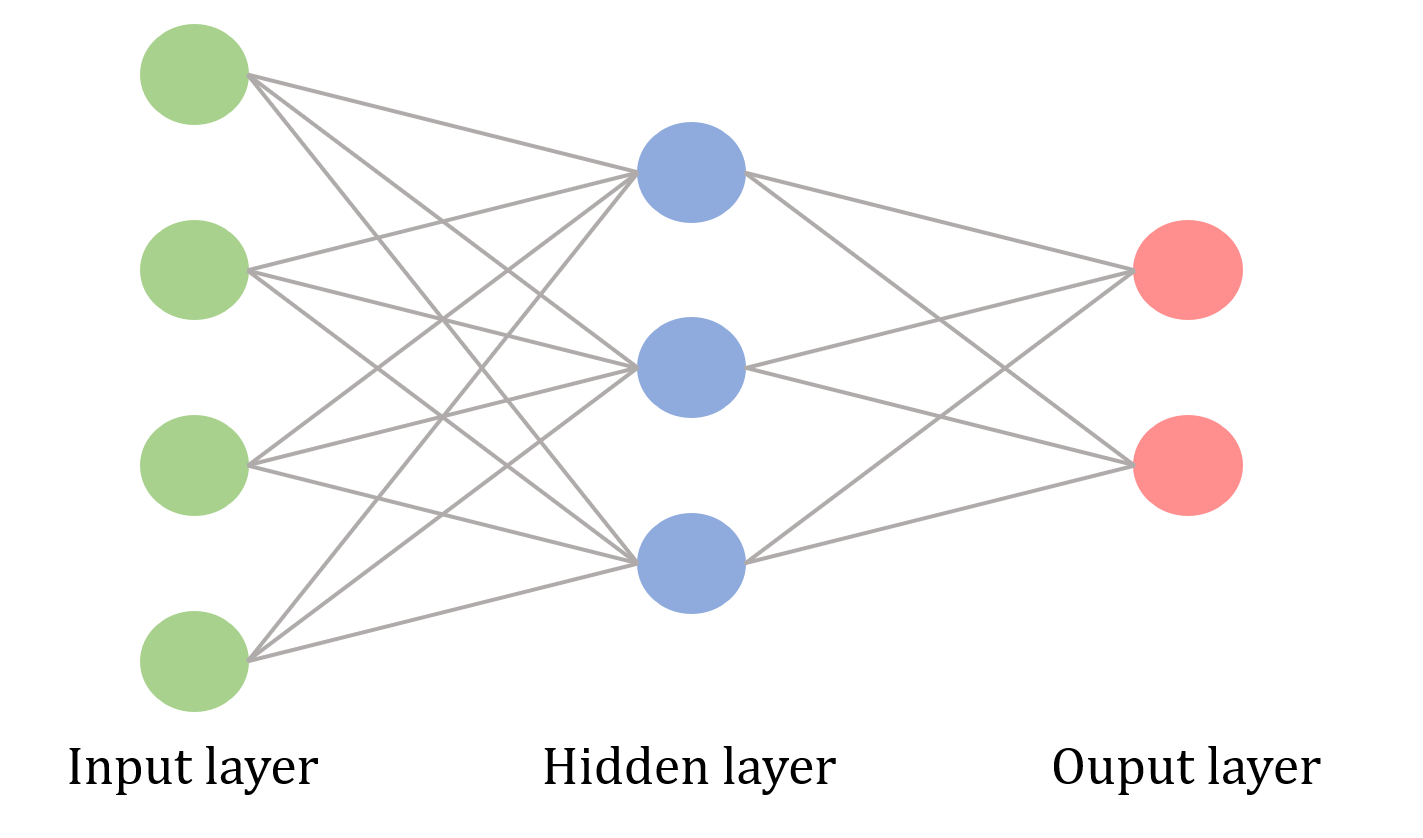}
    \caption{Fully-connected FNN architecture representation, with fully connected hidden layers enabling feature extraction.}
    \label{fig:dnn}
\end{figure}

A fundamental DL model can be represented as the fully connected Feedforward NN (FNN), a multiple-layered architecture, composed of fully-connected hidden layers. This model adopts a multi-layered architecture, with fully-connected hidden layers where every neuron is connected to all neurons in the adjacent layer, as illustrated in Figure \ref{fig:dnn}. Moreover, each of the layers corresponds to the representation of a distinct feature in the input data \cite{dl_phy_comm}, which highlights the model's enhanced ability to abstract information. The versatility of DNNs is a result of the flexibility in adjusting the number of neurons, layers, and the nature of connections based on the task the DL model is designed to  perform \cite{dl_phy_5g_6g}. Therefore, this section aims to depict some of the most relevant DL-based architectures and their applications and impact in communication systems.

\subsection{Semantic Extraction}
Convolutional NNs (CNNs) \cite{cnn} is an architecture specialized  designed for processing data with a structured grid-like topology \cite{deep_learning}. In CNNs, neurons are organized in a matrix form, creating feature maps that correspond to different data features. These models are distinct for having convolutional layers in the network, responsible for executing the mathematical operation of convolution on the input data \cite{deep_learning}. Given the challenges associated with the increasing number of parameters introduced by fully-connected FNNs, CNNs come as an attempt to address this issue by removing the connections between some of the neurons and their adjacent layers \cite{dl_phy_comm, dl_wireless}, resulting in localized connections \cite{dl_phy_5g_6g}. CNNs are a very relevant tool for image processing, given their capability to abstract and recognize features. 

Recurrent Neural Networks (RNNs) \cite{rnn} also come as relevant architectures for the deployment of Semantic Communications systems, tailored for processing sequential data, offering models a memory element that considers both current and past input data to compute outputs \cite{dl_wireless}. This functionality is realized through intra-layer connections, introducing a feedback loop system within the hidden layers \cite{dl_phy_5g_6g}. Moreover, RNNs are able to scale and handle varying data sizes. This design allows RNNs to capture temporal dependencies, where the state representations become a function of both the current input and the preceding hidden state. Nevertheless, it is important to refer that the basic or ``vanilla" RNN architecture does not enable contextual or sequential abstraction from prior states, limiting their performance with data with long-term dependencies \cite{dl_phy_5g_6g}.  

``Vanilla" RNNs face issues related to exploding or vanishing gradients when unrolled to a large length, impairing the proper learning process. Thus, the Long Short-Term Memory (LSTM) \cite{lstm} architecture was proposed as a type of RNN. LSTMs are able to capture longer-term dependencies in sequential data in comparison to the ``vanilla" RNNs, given their data flow control by introducing self-loops that add internal recurrence \cite{dl_phy_5g_6g, deep_learning}. Moreover, contextual abstraction is enabled by gated self-loops, where the definition of the weights is conditioned on the context and controlled by another hidden unit \cite{deep_learning}. Although LSTMs reduce vanishing and exploding gradients issues through the implementation of additive changes instead of multiplicative changes to update the states, this comes at a high cost, given the increased model complexity. Thus, one of the main advantages of LSTMs is their capability of grasping contextual information, which is very relevant in the context of Semantic Communications.

\begin{figure}[ht]
    \centering
    \includegraphics[width=6.5cm]{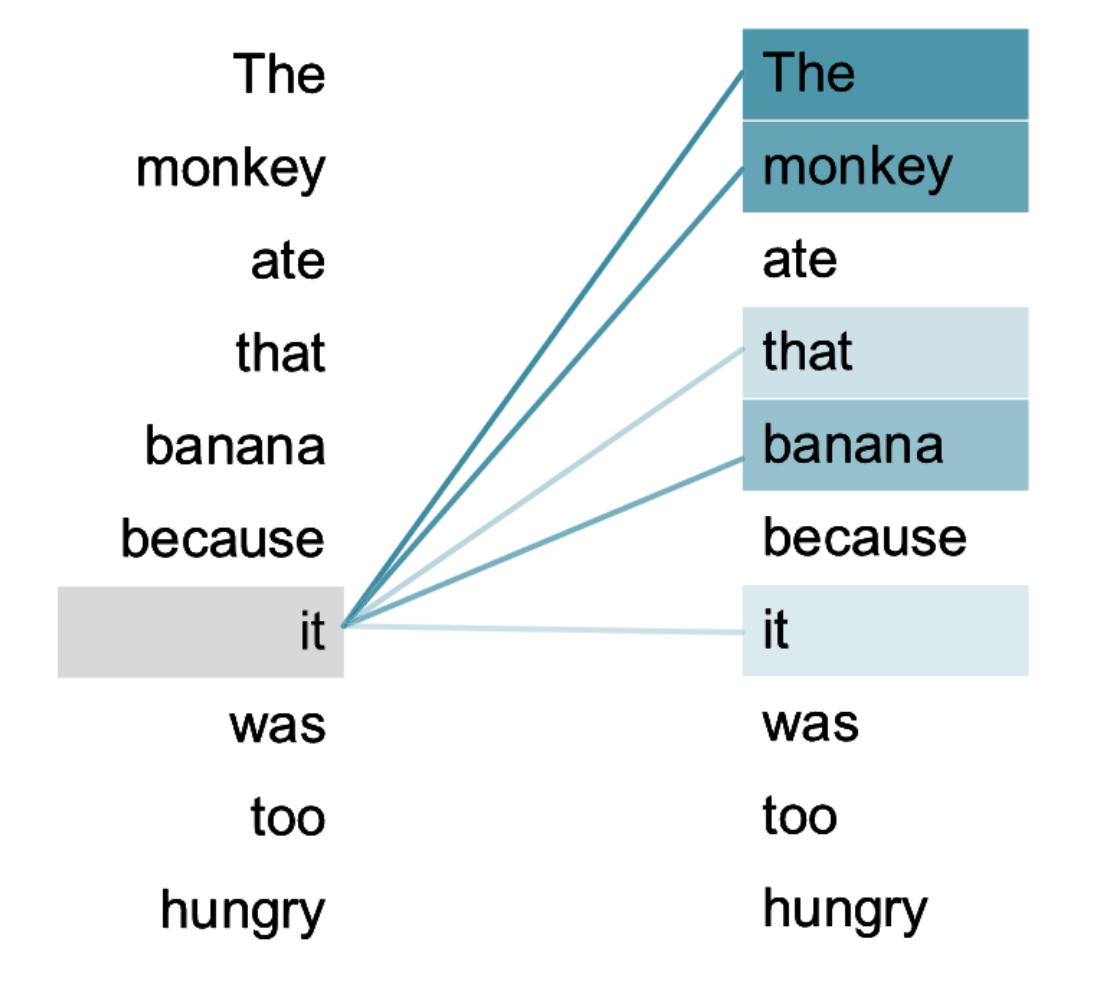}
    \caption[]{Self-attention mechanism visualization for long-term dependencies in the Transformer encoder \cite{deepsc}.}
    \label{fig:attention}
\end{figure}

In light of the challenges associated with RNNs in modelling long-term dependencies, Transformers, proposed in \cite{transformers}, revolutionized DL feature abstraction capabilities. The idea behind Transformers is that they rely solely on the attention mechanism to capture long-term dependencies, eliminating the constraints imposed by distance limitations inherent in recurrence and convolutions used in prior approaches \cite{transformers}. With an encoder-decoder stack architecture, Transformers are very flexible and scalable. As depicted in Figure \ref{fig:attention}, the attention  mechanism implemented in the Transformer encoder is able to abstract the contextual relationships behind the information, a valuable property in the implementation of feature extraction for semantic coding and decoding processes in Semantic Communications systems \cite{harq}.

\subsection{Deep Learning in End-to-End Communication Systems}
ML is envisioned to be a backbone of B5G technologies \cite{ml_5g, ai_enabled, ai_native, dl_wireless}. Traditional ML approaches, such as Support Vector Machine, Naïve Bayes, K-nearest neighbours, and logistic regression, have already demonstrated significant efficiency in enhancing various tasks within communication systems \cite{spectrum_sensing}. Tasks such as dynamic spectrum management \cite{knn_spectrum_auction, cr_dynamic_spectrum_access}, spectrum sensing \cite{spectrum_sensing, ml_spectrum_sensing}, and spectrum prediction \cite{spectrum_prediction}, demodulation and modulation recognition \cite{intro_phy} have all benefited from the integration of ML techniques. Nevertheless, with the evolution of more robust DL techniques, traditional ML-based systems have given way to DL-based counterparts. 

The utilization of DL-based optimizations in communication systems proves advantageous as it can identify imperfections and non-linearities inherent in practical systems. Due to DL being a highly adaptive approach, it is able to enhance system performance by considering specific hardware configurations and channel conditions \cite{intro_phy, dl_phy_comm}. Furthermore, NNs, as universal function approximators, exhibit a relevant capacity to model various systems, including the intricacies associated with challenging channel conditions within the Physical Layer \cite{intro_phy, dl_wireless}. In contrast, the fixed and constraining modular structure, with blocks for modulation, source encoding, and channel encoding at both the receiver and transmitter sides, impairs global system optimization \cite{intro_phy, dl_comm_review, dl_phy_comm}.

These suboptimal optimization results can be evidenced by the performance advantages for source encoding at lower data rates, whereas channel encoding tends to benefit from higher data rates \cite{jscc_wireless}. In light of this effect, Joint Source-Channel Coding (JSCC) schemes have already been investigated for traditional communication systems to enable End-to-End (E2E) optimizations \cite{jscc_wireless, lossy_jscc, jscc_codes, jscc_deep_space}. 

The distinctive characteristics of DL approaches enable their seamless integration into the Physical Layer, previously not possible with traditional ML techniques \cite{dl_phy_5g_6g, dl_comm_review}. By emulating E2E communication systems, DL demonstrates the capability to enhance the overall performance of communication systems. Hence, there is the potential to train the DNN as an Autoencoder (AE), where the physical channel is represented as a layer within the network architecture \cite{intro_phy}, minimizing a loss function defined according to the systems' objectives.

An AE refers to an unsupervised learning algorithm \cite{dl_phy_5g_6g}. With encoder and decoder layers, it can be used to replicate the architecture of communication systems, as first suggested by \cite{intro_phy}. In this context, the encoder layers are responsible for generating a low-dimensional representation of the input data through non-linear compression. The information is then reconstructed by the decoder layers with the highest fidelity possible. 

\begin{figure}[ht]
    \centering
    \includegraphics[width=0.5\textwidth]{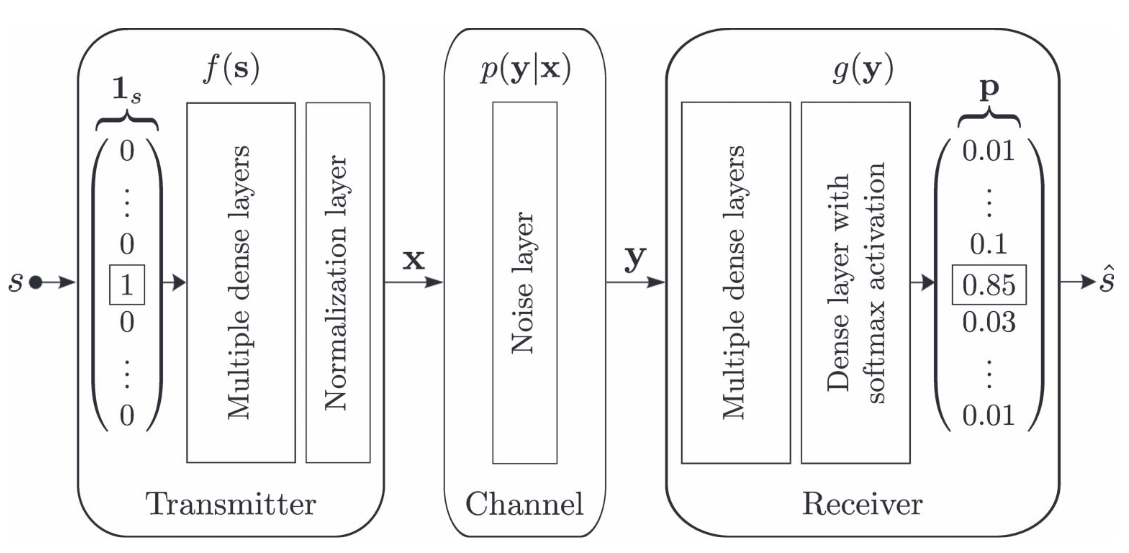}
    \caption{Autoencoder model for E2E communication system representation \cite{intro_phy}.}
    \label{fig:autoen}
\end{figure}

Hence, it is possible to say that AEs have the same objective as the communication problem defined by Shannon in \cite{shannon}: to reproduce, on one side, the closest representation possible of what was sent on the other. With the analogy between the encoder and decoder layers with the receiver and transmitter of communication systems, it is possible to use AEs to deploy E2E communication system optimization \cite{intro_phy, dl_phy_5g_6g}. In this context, the system is globally optimized to minimize a loss function. This loss function may refer to a network performance metric, such as the block error rate. This implementation is illustrated in Figure \ref{fig:autoen}.

In light of the time-sensitive nature of the intended applications for B5G technologies, JSCC is proposed as a viable alternative for enhancing information transmission efficiency and time consumption \cite{jscc_6g, deepjscc}. While JSCC has demonstrated efficacy in enhancing communication performance across various scenarios, its implementation complexity has hindered its widespread adoption \cite{deep_jscc_multiple_access}. Nevertheless, in recent times, the advancements in DL have led to effective and simpler implementations of JSCC systems, thereby allowing for the implementation of previously impractical systems. Therefore, DL emerges as a source of pertinent solutions for JSCC system developments, with the potential to offer universal, channel-agnostic solutions \cite{dl_phy_comm, dl_comm_review}. Additionally, one relevant advantage of DL-based implementations is that, even tho their training process is time-consuming, their parallelism enables significant improvements compared to the concatenation of state-of-the-art compression and capacity-approaching channel coding algorithms \cite{deep_jscc_image, deepjscc}.

\begin{figure}[ht]
    \centering
    \includegraphics[width=0.5\textwidth]{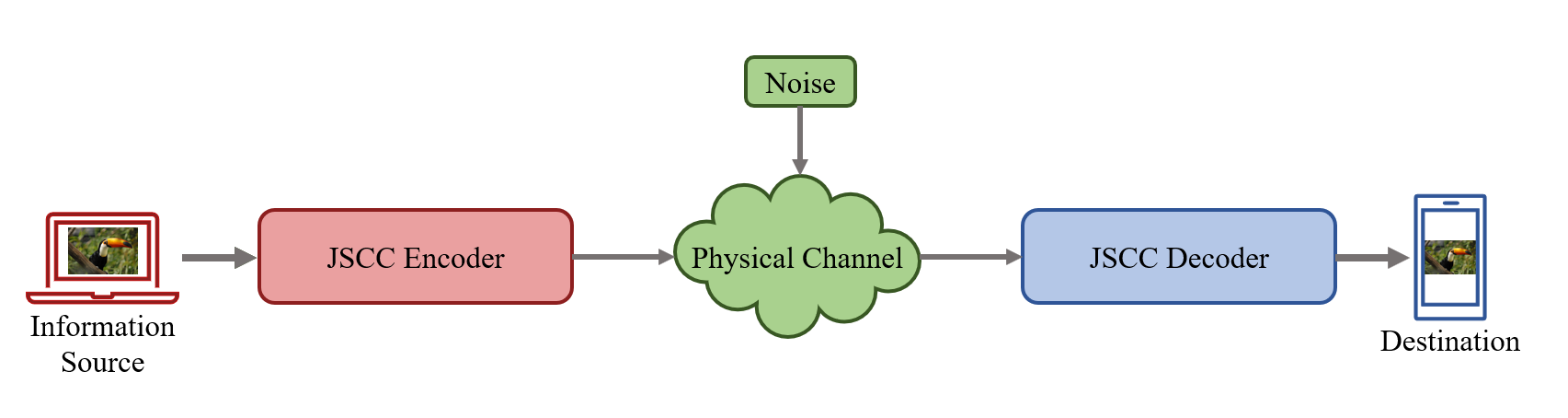}
    \caption{JSCC in Semantic Communications systems.}
    \label{fig:jscc}
\end{figure}

In the context of Semantic Communications, JSCC is regarded as an enabler approach to the realization of semantic encoding \cite{survey_on_sc}. Given that JSCC is based on the direct mapping of source signals to channel signal, allowing the receiver to infer the information estimations from the channel output, the communication system no longer relies on bits as the fundamental unit in the information transmission \cite{deepjscc}. Such a structure is intrinsically related to the implementation of Semantic Communications systems, aiming to minimize distortions in semantic information and transmit message features through transformations of the raw data into representations that are robust to the noisy channel interferences. Hence, DJSCC systems, firstly proposed in \cite{deep_jscc_image}, allow the joint optimization of the system through a data-driven approach, relying on DL models to simulate the communication process. Thus, it is possible to reduce the Semantic Communications system architecture presented in Figure \ref{fig:sc} to the one presented in Figure \ref{fig:jscc}. This allows the system to jointly learn optimal encoding and decoding strategies to improve the E2E system performance. Nevertheless, although vastly used, DJSCC is not the only approach for the implementation of Semantic Communications systems \cite{deepjscc}.

\begin{figure}[ht]
    \centering
    \includegraphics[width=0.5\textwidth]{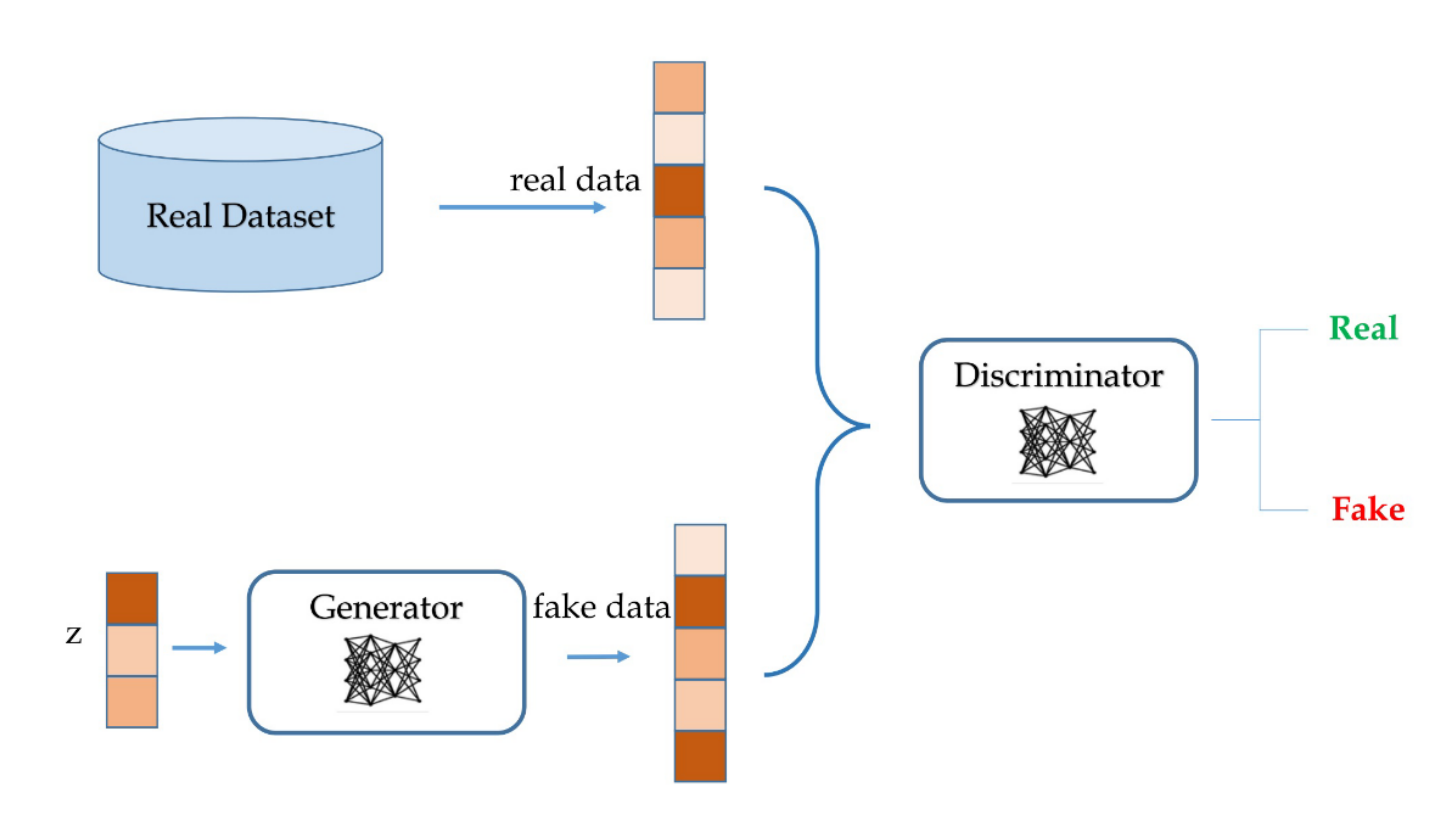}
    \caption{GANs model for E2E communication system representation \cite{gans_channel_agnostic}.}
    \label{fig:gans}
\end{figure}

However effective, this method imposes a restriction regarding the need to know the channel parameters for the system design, impairing the gradients flow through unknown channels \cite{dl_phy_comm, gans_channel_agnostic}. Given the impracticality associated with obtaining an accurate channel transfer function \cite{dl_phy_comm}, attempts to overcome this challenge have been made. In this context, another approach enabling DJSCC is the Generative Adversarial Network (GAN) architecture \cite{gans}. GANs come as another relevant approach in the deployment of E2E communications systems. The implementation of GANs is based on simultaneously training two models: the generative model G and the discriminative model D. While G is responsible for capturing the data distribution, D estimates the probability of a given sample being from the training data or generated by G \cite{gans}, meaning that it distinguishes ``real" and ``fake", as illustrated in Figure \ref{fig:gans}.

To simulate a communication system, the channel generator, transmitter and receiver are trained iteratively, in a process that allows the transmitter's parameters to be adapted through backpropagation to improve the quality of the generated samples. The training objective is for G to fool D, until it can only randomly guess whether the samples are real or generated \cite{gans_channel_agnostic}. Thus, GANs can represent the communication systems by learning to produce samples close to the distribution of the channel output, allowing the development of DJSCC systems that simulate the transmission under conditions that may be unknown or not analytically representable, like random effects such as channel noise and time-varying \cite{dl_phy_comm}.

\section{Implementations of Semantic Communications Systems}\label{sec:4}
As aforementioned, DL comes as a core feature in enabling Semantic Communications systems. With powerful abstraction capabilities that allow for both feature extraction and information reconstruction, it is possible to train the transmitter and receiver individually or to train the system in an E2E approach, as in DJSCC schemes. Additionally, Semantic Communications systems can also rely on embedding DL-based systems in traditional communication systems, increasing the applicability of the designed structure. This section presents implementations of Semantic Communications systems, classified here as either a Semantic Communications system design or real-world demonstrations or applications. 

\subsection{Semantic Communications System Designs}
According to the type of information transmitted, some relevant Semantic Communications systems designs are presented below. 

\subsubsection{Text Transmission}
With the aim of designing a DL-based Semantic Communications system for text transmission, \cite{deepsc} DeepSC, as one of the first DJSCC models to take advantage of the powerful abstraction capabilities of Transformers for semantic encoding and decoding. By considering the \textit{Level A} and \textit{Level B} of the communication problem, the proposed system was designed to allow intelligent E2E communication for variable-length sentences, while ensuring the robustness to semantic errors and improving the efficiency of the symbols' transmission. Thus, DeepSC system was optimized according to a loss function with two terms: one referring to maximize received and transmitted sentences similarity, and another based on the mutual information to maximize the data rate. One of the pioneering aspects of DeepSC is the use of BERT-based \cite{bert} semantic similarity measures to determine the semantic information recovery accuracy. The performance comparison of DeepSc revealed its robustness to low SNR values over the AWGN and outperformed all traditional communication baselines over Rayleigh channel, showcasing its ability to recover semantic information from distorted messages. Another advantage of the DeepSC is the reduced runtime in comparison to traditional communication systems, which is very relevant for low-latency transmissions. Moreover, an additional scenario with transfer learning for challenging environments was considered, showing that using the pre-trained model is not only more efficient but also led to better performance results.

Considering the resources constraints of Internet of Things (IoT) devices, \cite{l_deepsc} proposed the L-DeepSC as a lite distributed version of the DeepSC model. To ensure that the model is generalizable to diverse channel conditions, the L-DeepSC includes an attention-guided denoising convolutional neural network (ADNet) \cite{tian2020attention} to refine CSI, allowing improving the noise extraction from the received semantic features. Due to the computational limitations, DL models are commonly trained in edge/cloud platforms and then distributed to the IoT devices, being it imperative to reduce the number of parameters of the models. Thus, the proposed solution implements the compression of the DeepSC model by network sparsitication and quantization, reducing the model's size by pruning redundant neural connections and reducing the numerical precision weights. Moreover, to address the capacity constraints, the constellations learned by DeepSC are converted into finite-bits low-resolution constellation, enabling the implementation of L-DeepSC models in existing radio frequency systems. Over Rayleigh and Rician fading channels, both DeepSC and L-DeepSC outperform the considered traditional communication systems. Moreover, the L-DeepSC is shown to be a promising approach for IoT networks, achieving a similar BLEU performance to DeepSC, while having a reduced bandwidth usage and a lower runtime.

\begin{figure}[ht]
    \centering
    \includegraphics[width=0.5\textwidth]{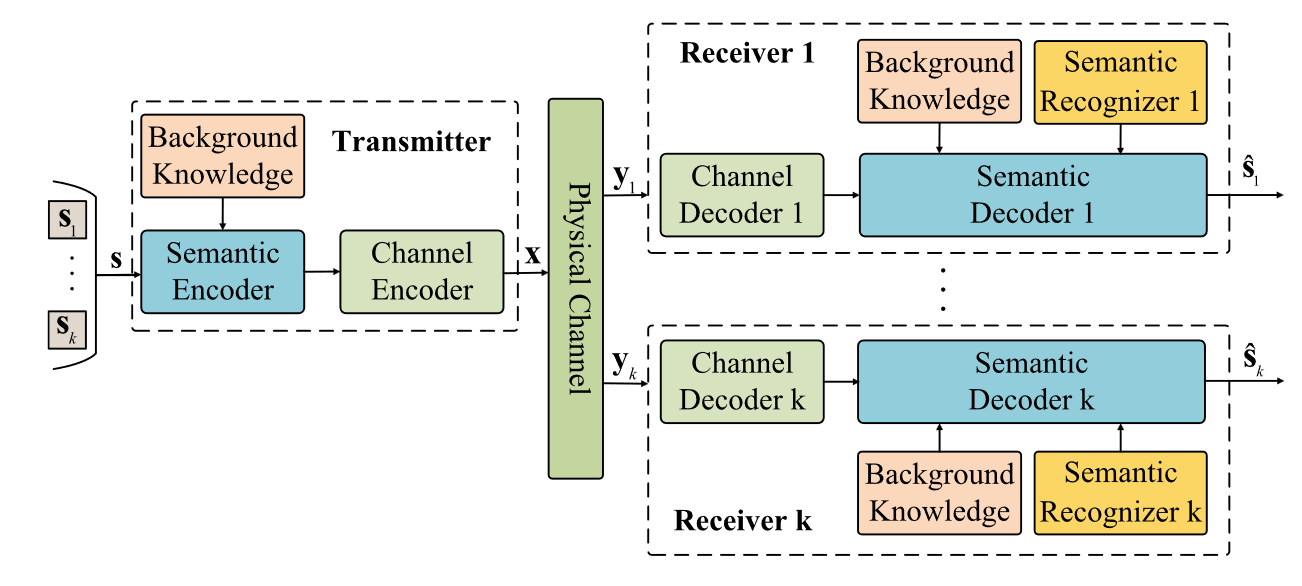}
    \caption{Framework of the one-to-many Semantic Communications system \cite{one_2_many_sc}.}
    \label{fig:one_to_many}
\end{figure}

Given the lack of analysis of broadcasting applications in Semantic Communications systems in previous studies, the work developed in \cite{one_2_many_sc} proposes the MR\_DeepSC one-to-many Semantic Communications system, represented in Figure \ref{fig:one_to_many}. Based on Transformers, the DJSCC scheme is designed to enable semantic information extraction and recovery, while the pre-trained DistilBERT model is used for semantic recognition. Traditional communication systems rely on division mechanisms to identify the users. In the MR\_DeepSC system, considering that different users should be assigned to a specific semantic features, the system relies on semantic recognition to distinguish receivers. Thus, this task can be seen as a semantic classification, associating features such as negativity and positivity to different users. To ensure the scalability and efficiency of the system, transfer learning is used for the addition of a new receiver network. Therefore, the training process relies on training the whole DJSCC model between the transmitter and one receiver, redesigning and training the semantic and channel decoders when adding a new receiver, freezing the parameters of the semantic and channel encoders. Through the BLEU performance evaluation, it is possible to conclude that the proposed solution outperforms traditional communication systems baselines for low SNR regime, increasing the robustness to channel conditions. Moreover, the unique contribution of this approach is that it can scale to more users without significantly degrade the system's performance.

Considering the lack of practicality associated with updating devices to support DJSCC-based Semantic Communications systems, \cite{semantic_importance} proposes Semantic Importance-Aware Communication (SIAC), a scheme that aims to use pre-trained Language Models (LMs) to integrate traditional communication systems through a cross-layer design. Thus, this approach focuses on assigning resource allocation priorities according to the semantic relevance of the frames, attempting to improve the reliable transmission under limited network resources by minimizing semantic loss. Here, the importance of a frame is then determined by the number of relevant words in the frame. In this work, the SIAC system tries to minimize the power consumption, considering implementations based on both ChatGPT and BERT. The performance is evaluated according to the expected important word errors and the expected semantic loss, both validating the proposed SIAC schemes in achieving lower power consumption in comparison to equal-priority transmissions.


\subsubsection{Speech Transmission}
When it comes to the high-fidelity reconstruction of raw speech signals through a DL-enabled Semantic Communications system, \cite{deepsc_s} proposed the DeepSC-S. Aiming to provide a general model of general model of Semantic Communications for the transmission of speech. Implementing a DJSCC system, the DeepSC-S aims to perform semantic extraction from speech signals and improve the reliability of transmissions over wireless channels. Thus, the deployment of DeepSC-S relies on the squeeze-and-excitation SE-ResNet attention mechanism for identifying the most relevant parts of speech according to the signal magnitudes and extracting the information from the speech signal. To determine the training conditions that would lead to a model generalizable to different channel conditions, the training under AWGN, Rayleigh and Rician channels was compared. It was concluded that the DeepSC-S trained over the Rician fading channel ensures a robust model, able to adapt well to other channel conditions. The SDR and PESQ performance evaluation show that DeepSC-S outperforms baseline solutions in all channel conditions, especially in the low SNR regime. Furthermore, one of the main advantages of DeepSC-S is the capability to provide reliable transmission over fading channels. The tests performed under telephone systems and multimedia transmission systems demonstrate the effectiveness of DeepSC-S, being it regarded as a promising candidate for speech transmission Semantic Communications systems.

\subsubsection{Image and Video Transmission}
\begin{figure}[ht]
    \centering
    \includegraphics[width=0.5\textwidth]{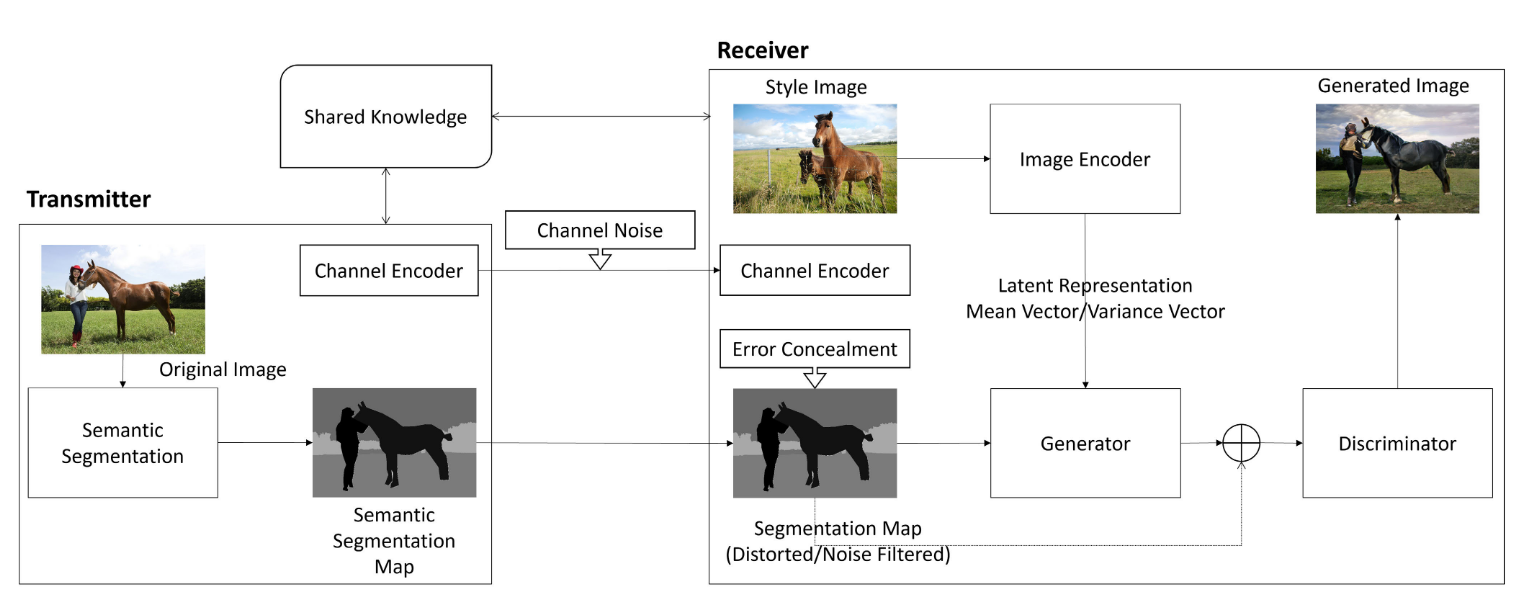}
    \caption{Semantic Communications system based on GANs \cite{wireless_e2e}.}
    \label{fig:wireless_e2e}
\end{figure}

The Semantic Communications system proposed in \cite{wireless_e2e} focussed on the optimization of bandwidth utilization for image transmission over a mobile communication channel. This system relies on the extraction of the segmented semantic map of the ground truth image at the transmitter. Upon channel encoding, the information is sent over a noisy channel to the receiver, which then proceeds with the channel decoding and uses the semantic segmentation to reconstruct the image with a pre-trained conditional GAN, as represented in Figure \ref{fig:wireless_e2e}. It is relevant to refer that this structure allows for the Semantic Communications system to have separated application, semantic and physical layers. The evaluation over AWGN channel, using polar codes for the channel encoding, showed that the proposed approach outperforms traditional JPEG compressed images transmission for low bit rates and SNRs, improving the robustness of the transmission. When it comes to the bandwidth optimization, the proposed approach was able to reduce the resource utilization by 95\%. Moreover, one relevant observation from this study is that, given that in semantic segmentation differences in pixel values refer to the edges of objects, object edge distortion was seen as a main cause of reconstruction degradation. In this work, the proposed solution to this issue was to use median filtering to remove salt and pepper noise to preserve the edges in semantic maps. Finally, the proposed system was also concluded to be possibly scalable for video transmission, highlighting the relevance and flexibility of this implementation.

Due to the stringent requirements of communication over Internet of Vehicles, with rigorous security needs and massive data sharing, \cite{segmentation} explored the efficient and reliable image transmission between vehicles. In this context, the Image Segmentation Semantic Communications (ISSC) is proposed, based on the accurate reconstruction of the semantic segmentation of images. This system is implemented through a DJSCC approach, with a cascade structure of Swin Transformer \cite{swin} layers to extract multiscale semantic features at the encoder. In turn, the decoder is composed of convolutional layers aiming to reconstruct the image segmentation, categorizing the image pixels into the different objects in the image. The performance evaluation of the system regarding the mean Intersection over Union (mIoU) reveals that, in comparison to traditional JPEG methods, the ISSC system was able to achieve a similar performance to the baseline under high SNR, with ISCC ensuring more stability and robustness under low SNR regime. Moreover, the proposed solution is also able to perform well under high compression rates. Thus, given the constraints associated with data transmission in vehicular scenarios, ISSC comes as a promising system, being able to reduce the amount of transmitted data given that it supports high compression rates.

\begin{figure}[ht]
    \centering
    \includegraphics[width=0.5\textwidth]{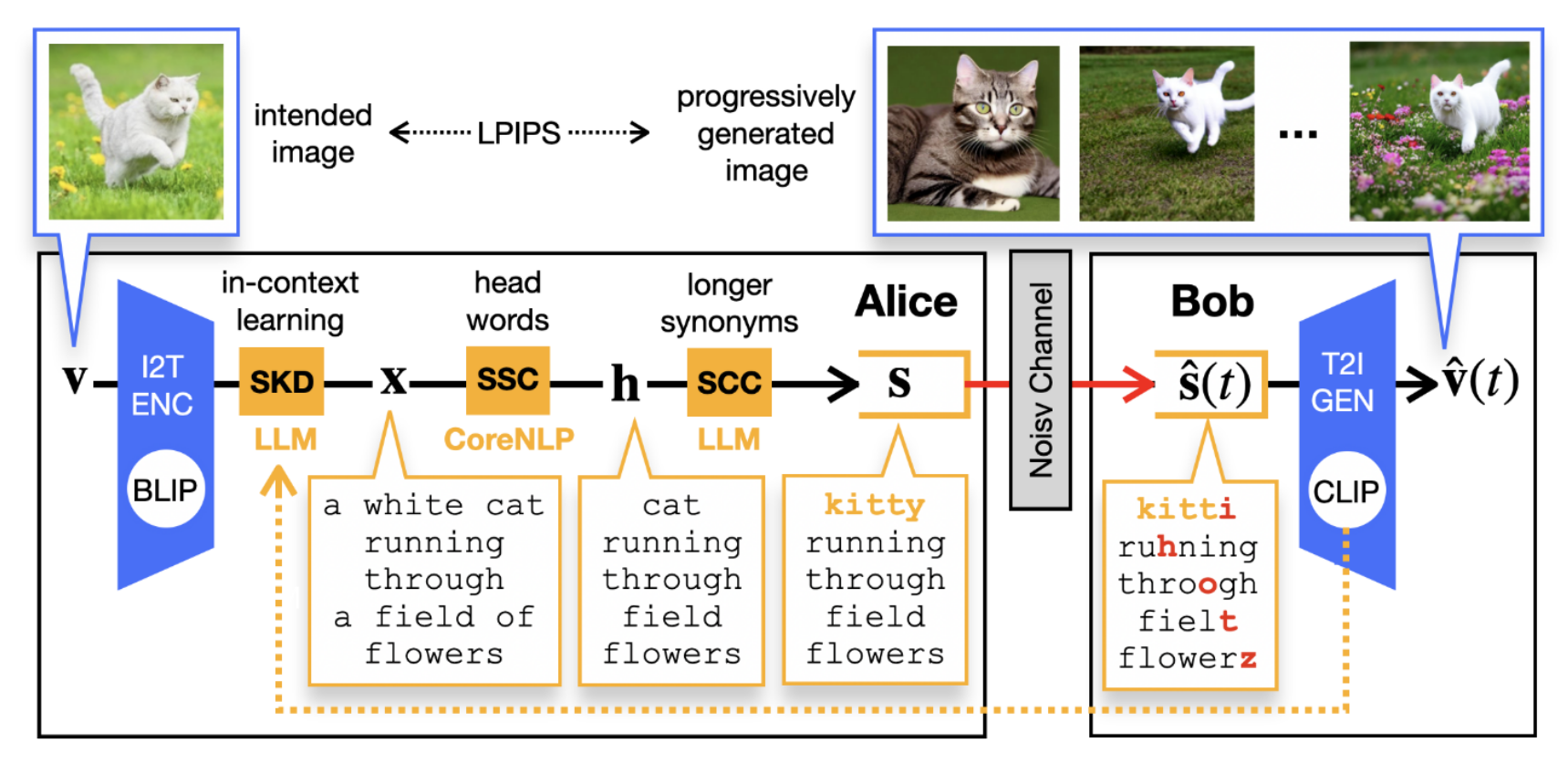}
    \caption{Language oriented Semantic Communications system framework \cite{text2image}.}
    \label{fig:clip}
\end{figure}

The work developed in \cite{text2image} proposes a language-oriented Semantic Communications system, as illustrated in Figure \ref{fig:clip}. This approach is based on the integration of LMs and generative models, with the transmitter relying on image-to-text encoders to generate a textual description of the image to be transferred. In turn, the receiver generates images progressively with a text-to-image generator, as it receives the words in the sentence describing the image to be reconstructed. This system relies on three algorithms to ensure the enhanced accuracy of received images: semantic source coding, semantic channel coding and semantic knowledge distillation. The semantic source coding is responsible for removing non-relevant words in the original description of the image, basically excluding stop words. Moreover, the semantic channel coding increases the transmission robustness by replacing small words by longer synonyms, to avoid misinterpretations. Finally, the semantic knowledge distillation is a method to avoid mismatches regarding the KB of the pre-trained LMs, specifying the desired context through few-shot learning. The performance was evaluated with the LPIPS and reveals the ability to increase the data compression, while proving the efficiency of the three proposed algorithms to improve the transmission quality.

In the context of image transmission for intelligent transportation systems, the lower latencies and reduced resource consumption of Semantic Communications system renders them as a useful tool. Thus, in \cite{nextg} a Semantic Communications framework for sequential images or videos transmission is proposed. This system comes as an attempt to address the limitation of previous works, such as \cite{wireless_e2e}, in achieving accurate identification of entities in the reconstructed images, through the extraction of semantic information at the transmitter side. This is because the focus is to reconstruct the objects in the image, without the need to ensure the recreation of the specificities of a given object, impairing, for instance, accurate identification of vehicles. The information is divided into static, referring to background information in the images, and dynamic, meaning the objects in the frame, which is the semantic information, delimited by a bounding box. In turn, at the receiver, a pre-trained conditional GAN for joint reconstruction and denoising of the received features. The performance evaluation reveals high sensitivity of the GAN to the SNR of the channel in the training process, which was overcome by using different SNRs during the training to improve generalizability. The system was shown to achieve more than 90\% reduction of the data transmitted through this periodic background information update system.

With the goal of analysing how to enable 360° constant video transmission to VR users in Semantic Communications systems, the work in \cite{wiser_vr} proposed the Wireless Semantic delivery for VR (WiserVR), with URLLC awareness. The main idea behind this implementation is to consider the coexistence of static, usually referring to the background, and dynamic objects in VR video frames. This separation allows the system to send the background information once and just transmit the updates on the objects, saving network resources. Moreover, this framework was implemented through a DJSCC scheme, with CNNs for image segmentation and Transformers performing video frame interpolation for behaviour recognition. The evaluation demonstrated the WiserVR substantially reduced both the latency and bandwidth consumption when compared to the traditional approach. Another relevant advantage of the WiserVR solution is the possibility of serving multiple VR users over multiple links. Nevertheless, although the results are promising, it is necessary to highlight that the evaluation was carried out for values of SNR between -9 and 6 dB, focusing on a much lower range than the other reviewed studies. 

Given the computation limitation of IoT devices, computational offloading to edge servers comes as a relevant solution to efficiently process information. To address the communication overhead and increase the efficiency of the transmission, in \cite{spatiotemporal}, a task-oriented Semantic Communications system was proposed for video action recognition. The designed Spatiotemporal Attention-based Autoencoder (STAE) aims to extract the semantically relevant frames and pixels per frame to reduce the transmitted symbols. The encoder is composed of frame attention modules, that select relevant frames for the final inference, and spatial attention modules, that determine the pixels that contain the most relevant information for the image classification task. In turn, the decoder is composed of a lightweight feature recovery module to reconstruct missing information, by abstracting temporal and spatial features. The evaluation of the proposed solution demonstrated that compression in the spatial dimension can maintain higher accuracy than compression in the frame dimension. Moreover, this implementation was faster and more efficient than the analysed baselines.

\subsubsection{Multimodal}
\begin{figure}[htb]
    \centering
    \includegraphics[width=0.5\textwidth]{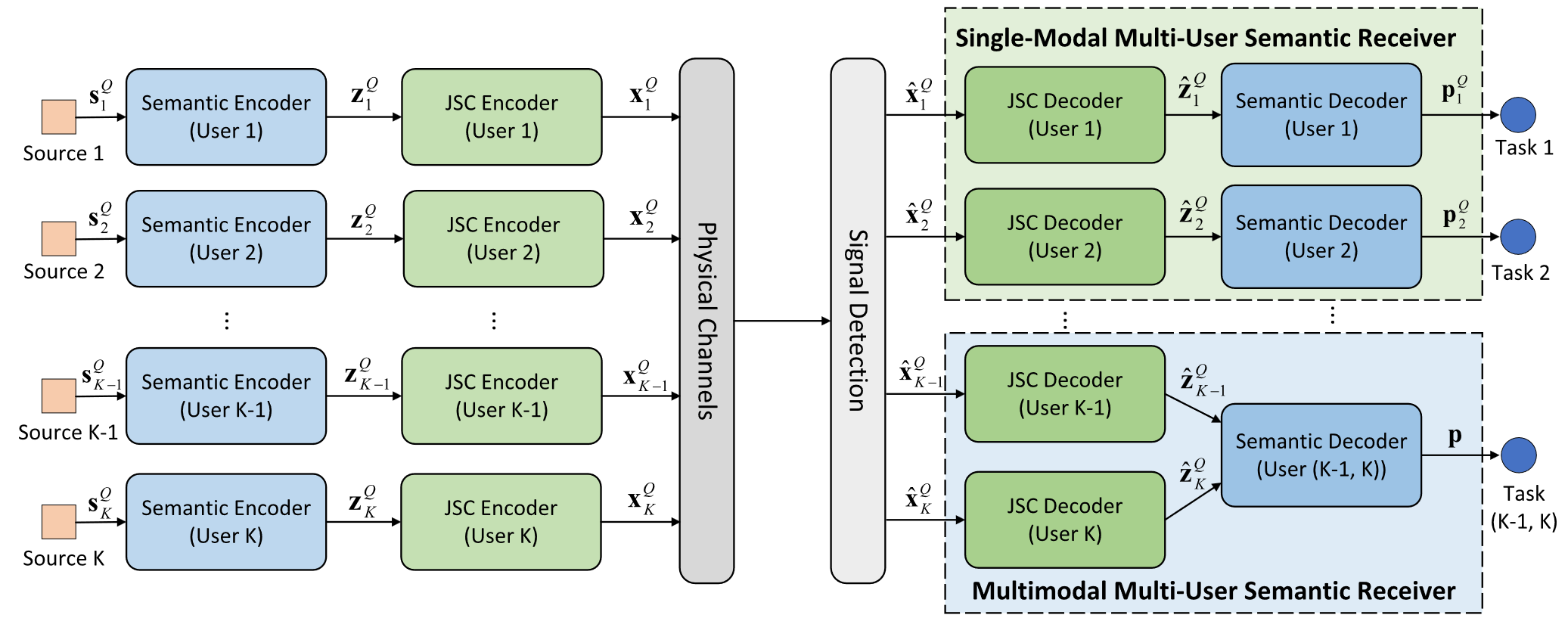}
    \caption{Task-oriented MIMO Semantic Communications system \cite{deepsc_vqa}.}
    \label{fig:vqa}
\end{figure}

Considering a MIMO scenario, the work developed in \cite{deepsc_vqa} presents a Semantic Communications system for single and multimodal data. With a unified DL-based Semantic Communications system, as represented in Figure \ref{fig:vqa}, three Transformers based transceivers were proposed: DeepSC-IR for image retrieval, DeepSC-MT for machine translation, and DeepSC-VQA for visual question answering. While both DeepSC-IR and DeepSC-MT are meant for single-modal transmission, of text and image, respectively, DeepSC-VQA is designed for multimodal data transmission of both images and text. To deal with the different data types, the system incorporates a layer-wise Transformer to facilitate the fusion of text and image information. The performance analysis of the proposed system shows that this solution was able to outperform traditional baselines, specially under low SNR domain. Furthermore, the framework exhibits the ability to transmit large data sizes with low power consumption. Specifically, both DeepSC-IR and DeepSC-VQA achieved a reduction of over 50\% in the number of transmitted symbols, leading to decreased bandwidth consumption and computational complexity.
    
In an attempt to design a Semantic Communications system that would enable the transmission of different types of data, \cite{multimodal} developed a Large AI Models-based Multimodal SC (LAM-MSC). To reduce resource consumption of the communication system and allow the transmission of various modal data through a unified model, LAM-MSC relies on converting audio, video and images into a text description of their content through the Composable Diffusion (CoDi) model \cite{codi}. Furthermore, a personalized large LM-based Knowledge Base (LKB) was proposed, for users to be able to have personalized prompts and fine-tune GPT-4 to improve semantic encoding and decoding tasks by reducing semantic ambiguity. Moreover, to enhance the system's performance over fading channels, a conditional GAN is employed to learn relationships between the received signal, the pilot sequence, and the CSI. The training process of the Semantic Communications model was based on a crossed-training strategy, alternating between the semantic and channel models. Finally, the performance evaluation validates the approach and proves the significant decrease on the bandwidth requirements, specially for video transmission.

\begin{figure}[ht]
    \centering
    \includegraphics[width=0.5\textwidth]{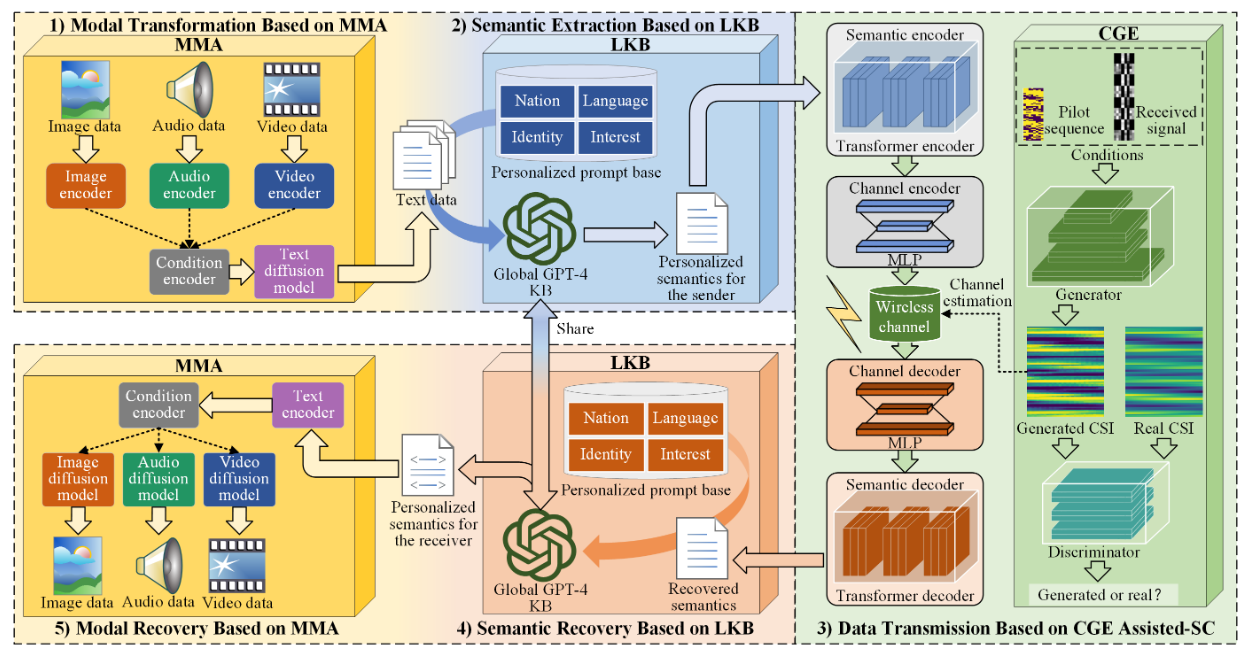}
    \caption{Multimodal LAM-MSC framework \cite{multimodal}.}
    \label{fig:multimodal}
\end{figure}

Motivated by the increased complexity introduced by the need for joint training in DJSCC-based Semantic Communications systems, the work carried out in \cite{genai_multimodal_prompts} refers to Generative AI methods as a potential solution. Given the randomness and instabilities associated with these approaches, the proposed system relies on multi-modal prompts for the accurate decoding of the received messages. Thus, associating visual and textual prompts, the receiver is able to leverage the image's structural information and the semantic description to reconstruct the original image with high fidelity. Additionally, this study also takes into consideration security aspects of the information exchange in wireless networks, mitigating risks of interference. The evaluation of this system focussed on the resource allocation optimization and lacks a more detailed evaluation of the proposed solution, establishing performance comparisons with other state-of-the-art Semantic Communications systems.

Considering the need to dynamically update information in the deployment of task-oriented Semantic Communications systems, the study conducted in \cite{editable_deepsc} proposes Editable-DeepSC, a cross-modal editable Semantic Communications system that uses StyleGAN \cite{stylegan} for adjusting images according to a textual prompt. Achieving significant compression rates, the Editable-DeepSC effectively reduces the communication overhead and increases fidelity, quality, and transmission efficiency, especially under noisy channel conditions. While data-oriented methods such as DeepJSCC \cite{deep_jscc_image} and traditional communications systems transmit and reconstruct the complete source information, the proposed solution allows transmitting only task-relevant information.

The comparison of the reviewed Semantic Communications systems are summarized in Table \ref{tab:papers1}, containing the aspects regarding the contribution, detailing the implemented architecture, specifying the wireless channel model used and the information object transmited, as well as the objective of the transmission. The solutions are organized according to the information object.

\begin{table*}[h!]
\centering
\caption{Summary of the reviewed Semantic Communications systems.}
\begin{tabularx}{\linewidth}{| >{\centering\arraybackslash}p{.035\linewidth} |  >{\centering\arraybackslash}p{.25\linewidth} |  >{\centering\arraybackslash}p{.15\linewidth} |  >{\centering\arraybackslash}p{.1\linewidth} | >{\centering\arraybackslash}p{.13\linewidth} |  >{\centering\arraybackslash}p{.19\linewidth} |}
 \hline
\textbf{Work} & \textbf{Contribution} & \textbf{Architecture}  & \textbf{Wireless Channel Model} & \textbf{Information object} & \textbf{Objective of Transmission} \\\hline 

\cite{deepsc} & Intelligent Semantic Communications system to maximize the system capacity and minimize the semantic errors. Demonstrated longer sentences improve reliability and is used as reference frequently  & DJSCC, with Transformer-based transceivers  & AWGN and Rayleigh fading channels & Text & Information recovery accuracy \\ \hline

\cite{l_deepsc} & Adapt the \cite{deepsc} model to a lite format, adjusting it for a distributed IoT system & DJSCC, compressed version of DeepSC \cite{deepsc} model & AWGN, Rayleigh and Rician fading channels & Text & Information recovery accuracy \\ \hline

\cite{one_2_many_sc} & First work to design one-to-many Semantic Communications system. Distinguish the intended receiver by message content (e.g., negative and positive) & DNN-based model, with Transformers for semantic extraction and recovery, and DistilBERT \cite{distilbert} for semantic recognition & AWGN and Rayleigh fading channels & Text & Information recovery accuracy \\ \hline

\cite{semantic_importance} & Reliable transmission of semantic meaning in resource-limited systems. Only introduces a cross-layer manager to traditional communication systems & Addition of cross-layer design to traditional communication systems & Rayleigh fading channel & Text & Resource allocation optimization \\ \hline

\cite{deepsc_s} & General model of Semantic Communications to reconstruct  speech sources. The attention module is able to extract relevant parts of the speech according to signal magnitudes &  DJSCC, with attention-based 2D CNN for channel and source encoding & AWGN, Rayleigh and Rician fading channels & Text & Information recovery accuracy \\ \hline

\cite{wireless_e2e} & GAN network at receiver to reconstruct received semantic segmentation into image, adding median filtering to enhance semantic decoding performance  & Pre-trained conditional GAN at receiver & AWGN channel & Image & Information recovery accuracy \\ \hline

\cite{segmentation} & First work to propose a Semantic Communications system to reconstruct image semantic segmentation & DJSCC, with Swin Transformers \cite{swin} & AWGN channel & Image & Task-oriented \\ \hline

\cite{text2image} & Language-oriented Semantic Communications system for the transmission of images with image-to-text and text-to-image conversion & Encoder with image-to-text encoder (e.g. BLIP \cite{blip}) and decoder with text-to-image generator (e.g. Stable Diffusion v1.5 \cite{stable_diff}) & AWGN channel & Image, converted to text for transmission & Information recovery accuracy \\ \hline

\cite{nextg} & Semantic Communications system for sequential images and videos transmissions in intelligent transportation systems & MSAM \cite{msam} at encoder and pix-2-pix conditional GAN \cite{pix2pix} at decoder
 & AWGN channel & Video and sequential images & Information recovery accuracy \\ \hline

\cite{wiser_vr} & First work to analyse VR applications in Semantic Communications systems. Delivers consecutive 360° video frames to users, with the awareness of URLLC & DJSCC, with CNNs for image segmentation and Transformers performing video frame interpolation & AWGN channel & Video & Information recovery accuracy \\ \hline

\cite{spatiotemporal} & Semantic Communications system for video action recognition & Attention-based Autoencoder & Time-varying Wireless Channel & Video & Task-oriented \\ \hline

\cite{deepsc_vqa} & Multi-user task-oriented approach, extracting and transmitting only the task-relevant semantic information. Supports both single-modal and multimodal data & DJSCC, with Transformer-based transceivers & AWGN and Rician fading channel & Multimodal & Task-oriented \\ \hline

\cite{multimodal} &  Convert audio, video and image into a text description, through which the receiver should be able to reconstruct original data & DJSCC, with CoDi, conditional GAN and ChatGPT & Fading channels & Multimodal, converted to text for transmission & Information recovery accuracy \\ \hline

\cite{genai_multimodal_prompts} & Optimization of resource allocation by using multi-modal prompts & Generative AI-based implementation with BLIP \cite{blip} & Wireless channel & Multimodal, sending both textual and visual prompts & Resource allocation \\ \hline

\cite{editable_deepsc} & Semantic Communications system focussed on the dynamic update of information & Generative AI-based implementation with StyleGAN \cite{stylegan} & AWGN channel & Cross-modal text-image & Task-oriented \\ \hline

\end{tabularx}
\label{tab:papers1}
\end{table*}

\subsection{Semantic Communications System Demonstrations}
To validate the possibility of real-world Semantic Communications systems deployments, demonstrations in simulation environments and simple testbeds are presented below, according to the type of information transmitted. Additionally, demonstrations in challenging environments are also reviewed.

\subsubsection{Image and Video Transmission}
The simulations carried out to analyse DJSCC methods do not account for effects as synchronization and non-linearities. To validate the promising capabilities of these approaches, the study conducted in \cite{real_time_sdr} an SDR-based platform for real wireless image transmission. The DJSCC model analysed in this study is the DeepJSCC system proposed in \cite{deep_jscc_image}, which is an AE-based model with CNNs. The implementation of this system includes GNURadio, to configure the transceiver, and Universal Software Radio Peripheral, for the RF communication system deployment. Thus, the DJSCC model, trained over AWGN channel, is applied through transfer learning to the GNURadio for real-time transmission. It is relevant to highlight that several adjustments were necessary to adapt the DJSCC model to the real implementation, given that DJSCC relies on directly converting the image features into channel symbols, while GNURadio are based on a modular digital processing approach. The evaluation of the real-time transmission shows transmission quality degradation under low SNRs in comparison with the software-based DJSCC implementation. due to the hardware and synchronization imperfections. Furthermore, the DJSCC system outperformed the traditional state-of-the-art baselines throughout the considered RF transmit gain interval. Thus, the DJSCC models are shown to have a graceful degradation even in real-time implementations.

To analyse real deployments of Semantic Communications systems for image and video transmission, in \cite{rosenfinch} the Rosefinch framework was proposed. The developed demo includes an extensive evaluation through real-time displaying of the performance, supported by a demo web page. Three scenes are studied in this context: medium-rate, low-rate and ultra-low-rate. The medium-rate scheme is explored for the case of video transmission, based on the concept of group of pictures, where the frames are predicted using an initial complete image. Moreover, the low-rate scheme was demonstrated for video conferencing, where the frames are reconstructed according to compact semantic information that serve as an update to an initial reference frame. This allows dynamic channel resources allocation according to the importance of the information. In turn, the ultra-low-rate scheme adopts a multi-modal approach that relies on the reconstruction of images according to their text description. The system was evaluated through the rate and distortion, revealing that, in comparison to traditional systems, Rosefinch was able to significantly reduce the payloads and rate, while also achieving stable transmissions with fewer distortions.

A demonstration of a E2E DL-based Semantic Communications system was detailed in \cite{real_time_vit}, aiming to prove its feasibility of image transmission through a field-programmable gate array. The encoder was implemented with convolutional layers and residual blocks, which improved the feature abstraction capabilities. On the other hand, the decoder relied on a vision transformer and a denoising AE to reduce the effect of noise in the received images. The model was trained under AWGN and Rayleigh channels and evaluated on a real-time wireless transmission system, with the results demonstrating that the proposed solution outperformed the traditional communication system used as baseline for the image reconstruction quality for low SNRs. 

Following the work presented in \cite{real_time_vit}, the analysis is extended to not only the application of vision transformers, but also the fine-tuning of the network to improve the system's architecture in \cite{ViT_sc_prototype}. Thus, this work aims to provide a detailed analysis of the functioning of the DJSCC implementations, using a model composed of layers of vision transformers and CNNs. This concatenation of different approaches to capture the images' features, taking also advantage of the capabilities of the CNNs in terms of channel coding, as well as of the vision transformers for source coding and increased robustness to noisy signals. With such a configuration, the proposed model is characterized by a significative content-adaptativeness, which enables it to reduce the redundancy and increase the diversity of the generated output features. Moreover, the developed solution was evaluated on a Software Defined Radio-based wireless Semantic Communications testbed, showing results that outperformed the state-of-the-art DeepJSCC model under all evaluated conditions. 

The demonstration carried out in \cite{talking_head} points to the promising performance of the proposed prototype for facial video reconstruction in video conferences. This system was designed to enable the encoder to perform key-point facial feature extraction and, from an initial reference frame, update facial expressions and eventual movements through a generative approach. Thus, with a parallel design, this system substantially reduces the reconstruction latency and the transmitted code rate.

\subsubsection{CSI Feedback}
Massive MIMO is one of the core features of 5G technologies, with B5G networks being expected to support even higher connectivity capabilities. However, given that MIMO technologies rely on the knowledge of the CSI transmitted to the base station, it is essential to ensure CSI feedback reconstruction under diverse channel conditions. This issue is addressed in \cite{deep_jscc_csi}, proposing the ADJSCC-CSI model, as a general DJSCC solution to modelling the CSI source and feedback channel accurately. Thus, ADJSCC-CSI comes as a general framework composed of three modules, the nonlinear transform module, which aims to reduce distortions in the CSI feedback compression in traditional communication systems, the DJSCC module, enabling the upgrade of existing CSI compression networks into a DJSCC-based model, and the SNR adaption module, with attention feature modules and responsible for allowing a single DJSCC network to be trained with various SNRs. Moreover, the system is modelled as a frequency-division duplexing massive MIMO orthogonal frequency division multiplexing system and evaluated in an open door scenario. The performance assessment reveals the relevance of the AS modules, ensuring the SNR adaptation, with the proposed solution outperforming the baselines under all scenarios. Furthermore, ADJSCC-CSI was able to effectively upgrade existing CSI solutions into a DJSCC model, which was also proven to be very beneficial for the system's performance. Moreover, the ADJSCC-CSI framework was able to effectively spare resources in both the user equipment and the base station.

\subsubsection{Challenging Environments}
Due to the underwater acoustic channel being characterized by uncertainty and low quality transmissions, adaptive communication schemes are desired to overcome the challenging and dynamic conditions. Thus, in \cite{underwater}, a DJSCC Semantic Communications data- and channel-aware system is designed for underwater image transmissions, relying on CSI feedback to adapt the information encoding. The feature extraction is executed with CNNs in an unsupervised approach, allowing an optimization of the identification of objects to compress the images according to the semantic content of the different regions. With a variable-length encoding design, one of the aspect of the implementation is that the encoded features and CSI information are submitted to LSTM-based DJSCC encoding, where the coding problem is implemented as a machine translation task through sequence-to-sequence model. The model is trained under a Rician fading channel, as it resembles the complex conditions of underwater transmissions. The proposed solution is evaluated with a pool testbed, however, the analysis focusing solely on network metrics and the quality of the received images was not explored, impairing efficacy examination.

In the context of non-terrestrial networks, Unmanned Aerial Vehicles (UAVs) can be used to sense spectrum availability in remote areas. Given the limited resources and low latency requirements, in \cite{sc_remote_sensing} a Semantic Communications system was proposed to enable real-time UAV-BS collaborative spectrum sensing, with highly efficient transmissions. In this context, a DJSCC-based framework was designed to allow the UAV to extract the relevant features from the statistical covariance matrix of the signals collected from the end user. The decoding is then done by the base station, allowing it to determine the status of the end user by the received features. This system is further integrated with a reinforcement learning approach for the UAV to adjust its trajectory according to the signals received from the end user. Through simulation assessments, the solution was shown to approach the ideal detection performance, while reducing the transmission latency.

Considering the challenges posed by their computational limitations, \cite{sc_iot} focused on the development of a Semantic Communications system that would allow efficient image communication in IoT systems for AI, with the objective of achieving accurate reconstruction of the original images. In this context, the designed transmitter performs high-precision semantic segmentation, while also ensuring compression to increase efficiency. In turn, the receiver aims to provide a high fidelity reconstruction of the input image through a cyclic GAN. Moreover, with residual blocks and skip connections, the decoder has access to detailed feature information, improving their abstraction capabilities. The performance evaluation shows the efficiency and low latency of the proposed model, outperforming traditional pixel-based methods for the compression of high dimensional data. This is because of the semantic segmentation approach being able to detect and adapt compression according to the different regions and objects in the image, maintaining the quality of the semantic information. Moreover, the analysis of both the PSNR and SSIM values indicate that the proposed system was able to ensure higher transmission quality than traditional systems. Finally, this study presents a real demonstration, however, the results seem to have been used just for the latency analysis, lacking also any examination of the system's behaviour under different channels or SNR conditions.

The comparison of the reviewed demonstrations of Semantic Communications systems are detailed in Table \ref{tab:papers}.
\begin{table*}[ht]
\centering
\caption{Demonstrations summarization.}
\begin{tabularx}{\linewidth}{| >{\centering\arraybackslash}p{.035\linewidth} |  >{\centering\arraybackslash}p{.25\linewidth} |  >{\centering\arraybackslash}p{.15\linewidth} |  >{\centering\arraybackslash}p{.1\linewidth} | >{\centering\arraybackslash}p{.13\linewidth} |  >{\centering\arraybackslash}p{.19\linewidth} |}
 \hline
\textbf{Work} & \textbf{Contribution} & \textbf{Architecture}  & \textbf{Wireless Channel Model} & \textbf{Information object} & \textbf{Objective of Transmission} \\\hline 

\cite{real_time_sdr} & First Software Defined Radio-based prototype of DJSCC wireless  image transmission & DeepJSCC \cite{deep_jscc_image} & AWGN channel (training) and air (demo) & Image & Information recovery accuracy \\ \hline

\cite{rosenfinch} & Prototype of Semantic Communications system for video conferencing and video and image transmission & Not detailed, but supposed DJSCC & Wireless channel (training) and web transfer (demo) & Image and video & Information recovery accuracy \\ \hline

\cite{real_time_vit} & First work to implement real-time Semantic Communications with a vision transformer & DJSCC, with ViT and denoising AE at the decoder & AWGN and Rayleigh channels (training) and air (demo) & Image & Information recovery accuracy \\ \hline

\cite{ViT_sc_prototype} & First analysis of fundamental functional aspects of a semantic communications system, also supported by a prototype & DJSCC, with ViT and CNNs & AWGN and Rayleigh channels (training) and air (demo) & Image & Information recovery accuracy \\ \hline

\cite{talking_head} & Ultra-low bit rate face video Semantic Communications system & Generative approach & Physical channel, not detailed (demo) & Video & Resource allocation optimization \\ \hline

\cite{deep_jscc_csi} & DJSCC system for CSI feedback and unified DJSCC module, allowing architectures of existing CSI compression networks to be upgraded to a DJSCC based network & DJSCC, with CNNs and attention feature modules & Fading channel (training) and air (demo) & CSI Feedback & Information recovery accuracy \\ \hline

\cite{sc_remote_sensing} & Application of Semantic Communications to remote sensing & DJSCC
 & AWGN channel (training) and simulation environment (demo) & Spectrum sensing & Task-oriented \\ \hline

\cite{underwater} & Model encoding as the translation of the input sequence to another ‘language’, depending on the estimated channel conditions & DJSCC, with CNNs & Rician fading channel (training) and water (demo) & Image & Information recovery accuracy \\ \hline

\cite{sc_iot} & Prototype of SC system for information compression and reconstruction accuracy, with delay evaluation & CNNs for feature extraction at encoder and cyclic GAN at decoder & AWGN and Rayleigh fading channel (training) and air (demo) & Image & Information recovery accuracy \\ \hline

\end{tabularx}
\label{tab:papers}
\end{table*}

\section{Open Challenges}\label{sec:5}
The transition towards a Semantic Communications paradigm brings significant advancements, introducing the design of high-performance and flexible systems that hold the promise of outperforming traditional communication systems across various scenarios \cite{deepjscc}. Despite the potential benefits, it is crucial to acknowledge that Semantic Communications systems are relatively recent, with only a few system demonstrations having been held so far. Therefore, there are still many aspects to be clarified and, thus, extensive research should still be conducted before their actual implementation. Some of the open challenges in Semantic Communications systems are detailed below. 

\subsection{Generalization}
One of the main issues come as the generalization challenges faced by Semantic Communications systems. Particularly, for the case of applying a DJSCC architecture, transceivers must be trained together, limiting the interoperability if separately trained. Moreover, the joint source and channel coding is based on the model being trained to map input information into I/Q samples for channel transmission. Therefore, the DJSCC models can only be used in point-to-point transmissions, which can impair their vast applicability with existing devices \cite{robust_sc_gan}.

Moreover, the implementation of Semantic Communications systems is highly dependent on the information object that is transmitted, which raises impairs related to multi and cross-modal system implementations. In this context, the performance of the system is determined from its ability to perform a specific task at the receiver from the received features or to reconstruct the original information from the features. Within the same information type, it is possible to use different metrics to determine the quality of the reconstructed information, illustrated by the possibility of using SSIM or PSNR for images. This highlights the lack of a unified measure to analyse the performance of Semantic Communications systems, allowing also fair comparison with other implementations. As referred in \cite{sc_principles_challenges}, it is necessary for Semantic Communications systems to have a general performance metric, such as the bit error rate in traditional systems. 

\subsection{Scalability}
When it comes to real-world deployments, challenges regarding the implementation of DJSCC schemes in proprietary traditional communication devices are detailed in \cite{underwater, real_time_sdr}. This is also shown by the fact that so far very few Semantic Communications systems demonstrations have been carried out, evidencing the need to either adapt the communication technologies to the semantic paradigm or develop Semantic Communications systems that can be deployed on traditional communication devices. Moreover, this highlights research opportunities not only for the design, but also for the implementation of Semantic Communications systems. 

Another scalability issue is the development of distributed multi-user systems, which requires not only recognizing the different physical signals according to the user, but also handling different KBs, which should be regularly updated both at the source and destination \cite{sc_overview, sc_survey_landscape}. Relying on KBs for the feature extraction and information reconstruction also reveals the need for additional storage and computational demands in the communicating devices. These challenges should be addressed to allow the development of complex multi-user Semantic Communications systems.

\subsection{Channel Modelling with Deep Learning}
Given the need to guarantee the flow of gradients through the network for the optimization through backpropagation, the channel layer implemented in the DL-based system should be differentiable. This limits the possible representation of diverse channel conditions for the training of the DNN, thus restricting its applicability. Additionally, in the context of emerging 6G services, it is essential that Semantic Communications systems are designed to ensure efficient channel allocation and increased robustness to varying channel conditions \cite{ml_5g}. 

One way to overcome this issue was pointed out in \cite{deepsc_s, underwater}, where it was demonstrated that training the model under the Rician fading channel improved their generalization capabilities, also enhancing the model's robustness to the low SNR regime. Moreover, as suggested in \cite{underwater}, the decoding of the received information can be seen as a machine translation task, which opens the possibility of resorting to LMs to perform the denoising of received codes. However, most Semantic Communications systems designs are still not channel agnostic, and further research is needed to ensure that models are generalizable to diverse channel conditions. 

\subsection{Overfitting and Underfitting}
\begin{figure}[ht]
    \centering
    \includegraphics[width=0.5\textwidth]{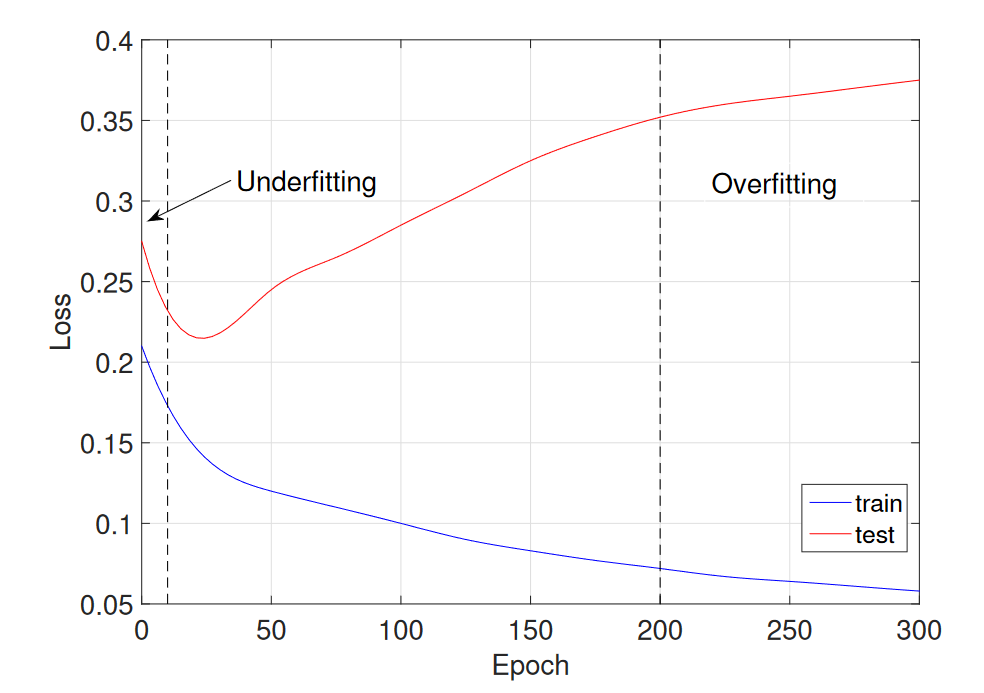}
    \caption{Underfitting and overfitting recognition in the training and test losses \cite{overfitting}.}
    \label{fig:under_over_fitting}
\end{figure}

As all ML systems, if not carefully implemented, DL-based E2E communication systems such as DJSCC schemes can suffer from underfitting and overfitting, impairing their reliable deployment \cite{overfitting}. Both effects can be spotted in the evolution of the training and testing losses, as illustrated in Figure \ref{fig:under_over_fitting}. The problem of overfitting comes from the lack of generalization of the trained model, showing good performance on the training data, while not adapting well for the test data. This is a result of overly complex NN architectures, unable to adapt to the characteristic distributions of the test data \cite{deep_learning}. 

As demonstrated in \cite{overfitting}, the DL-based E2E communication systems can benefit from regularization techniques, such as dropout and L2-norm regularization to avoid overfitting. In turn, underfitting refers to the inability of the DNN to perform well on the training data, as a result of the simplicity of its architecture. Thus, the model is unable to abstract complex characteristics of the data \cite{overfitting}. With the literature review, it was possible to see that many models faced overfitting for specific SNRs, impairing their generalization capabilities under diverse channel conditions. Nevertheless, the overfitting issue was not explicitly detailed in these studies, and it is imperative to conduct further in-depth analysis on overfitting and underfitting in E2E communication systems.

\subsection{Communication Overhead}
Although the Semantic Communications systems approach has shown to have promising results, there is a lack of analysis regarding the overhead introduced in the implementation of Semantic Communications, particularly for text transmission. While data compression is achieved and demonstrated for image and video transmission systems, the transmission of word or sentence embeddings, as high dimensional vectors, increase the amount of data transmitted in comparison to the original text. 

Furthermore, the additional training process required to developed DL-based systems is very costly, particularly if architectures are based on Transformers. Moreover, it is also necessary to ensure regular updates to the KB databases to all the receiver-transmitter in the network. This can be very challenging to achieve in large distributed systems. The KB update should be followed by a retraining or fine-tuning of the transceivers. This process faces many limitations in unrealistic practical systems, due to their limited computational resources \cite{sc_fundamentals}. Thus, this raises a trade-off between the system's accuracy and the computation overhead that should be the object of carefully evaluation.

\subsection{In-Depth Evaluation of the Semantic Communications systems}
The studies reviewed in this paper primarily concentrate on performance evaluations linked to objectives related to information transmission, such as resource allocation, accuracy of information retrieval, and performance in downstream tasks. However, there appears to be a gap in the analysis concerning network metrics, highlighting the need to provide detailed characterization of systems' overall resource consumption and potential communication overhead introduced.

\subsection{Security in Semantic Communications systems}
 In traditional communication systems, the encryption is performed after source coding and before channel coding. Thus, traditional encryption techniques are not compatible to DJSCC systems, as there is a direct mapping from the source information to channel input symbols \cite{deepjscc}. As proposed in \cite{deep_jscec}, public key encryption can be applied to DJSCC systems, encrypting transmitted symbols and training the network to increase the robustness against the additional encryption noise. However promising, it is imperative for researches to address this issue and improve the efficiency of encryption in DJSCC systems. 
 
Moreover, the paradigm shift to Semantic Communications highlights the emergence of the KB as an essential element, not only in the feature extraction process at the transmitter, but also for the reconstruction of the information at the receiver. This new approach opens the door for exploiting vulnerabilities regarding the KB databases and feature extraction process, which should also be carefully analysed to ensure security in Semantic Communications systems.

\section{Conclusions}\label{sec:6}
The ever-increasing complexity and stringent requirements that characterize the envisioned B5G network system and applications evidence the need to develop a communication system that allows to go beyond Shannon's Information Theory. In this context, Semantic Communications comes as a communication paradigm based on the transmission of ``semantic information", relying on recent advancements achieved with DL technologies to enable feature extraction and system optimization processes. 

The literature review carried out in this study encompassed the fundamentals of Semantic Communications systems, as well as an analysis of some relevant designs and demonstrations performed so far. Overall, in comparison to traditional communication systems, the semantic-oriented approaches enable more stable and robust deployments, more resistant to channel effects such as fading. However, it was possible to observe that the system performance is highly dependent on the type of information transmitted. For images, the main benefit is the compression enabled by the separation of background environment and object, which reduces the transmitted data. Similarly, for video, Semantic Communications systems reduce the bandwidth consumption by sending an initial image and just updating frame changes. Nevertheless, for text, the improved robustness comes at the cost of the resource consumption, which was not thoroughly analysed in the reviewed works and cane be higher than in traditional communication systems. As the implementation of Semantic Communications systems was enabled by DL, this approach is very recent, and many challenges should still be addressed before its wide adoption. Thus, the opportunities and research questions that arise within this theme motivate many detailed studies in the field.

\bibliographystyle{IEEEtran}
\bibliography{refs}
\end{document}